\documentclass[journal]{IEEEtran}

\usepackage{cite}
\usepackage{amsmath,amssymb,amsfonts}
\usepackage{algorithmic}
\usepackage{graphicx}
\usepackage{textcomp}

\usepackage{multirow}
\usepackage{subcaption}
\usepackage{units}

\DeclareCaptionSubType[alph]{figure}
\makeatletter
\renewcommand\p@subfigure{\thefigure~}
\makeatother

\usepackage{fancyhdr}
\usepackage{upgreek}
\usepackage{wrapfig}
\usepackage{mathtools}
\usepackage{comment} 
\usepackage[ansinew]{inputenc}
\usepackage[T1]{fontenc}
\usepackage[british]{babel}
\usepackage{csquotes}
\usepackage{hyperref}
\hypersetup{
    colorlinks,
    citecolor=black,
    filecolor=black,
    linkcolor=black,
    urlcolor=black
}
\usepackage{xcolor}
\def\BibTeX{{\rm B\kern-.05em{\sc i\kern-.025em b}\kern-.08em
    T\kern-.1667em\lower.7ex\hbox{E}\kern-.125emX}}

\definecolor{blue(pigment)}{rgb}{0.2, 0.2, 0.6}

\begin{document}
\urlstyle{tt}

\markboth{2020 Accepted in IEEE Transactions on Circuits and Systems-I Journal / DOI:10.1109/TCSI.2020.2981901\hfill}{}

\title{IMAC: In-memory multi-bit Multiplication and ACcumulation in 6T SRAM Array \\
}
\author{Mustafa Ali*,~\IEEEmembership{Student Member, IEEE}, Akhilesh Jaiswal*, Sangamesh Kodge*,  Amogh Agrawal,\\ Indranil Chakraborty and Kaushik~Roy,~\IEEEmembership{Fellow,~IEEE}
\\ (* Equal contributors listed in alphabetic order)
\thanks{
The authors are with the School of Electrical and Computer
Engineering, Purdue University, West Lafayette, IN 47906 USA 
(e-mail: \{ali102, jaiswal,skodge,agrawa64,ichakra,kaushik@purdue.edu\}). All the codes used in this work can be found on \url{https://github.com/Sangameshkodge/IMAC}.
\newline
\textbf{Acknowledgement:} This work was supported by C-BRIC, one of six centers in JUMP, a Semiconductor Research Corporation (SRC) program, sponsored by DARPA. 
}
}
\maketitle

\begin{abstract}
`In-memory computing' is being widely explored as a novel computing paradigm to mitigate the well known memory bottleneck. 
This emerging paradigm aims at embedding some aspects of computations inside the memory array, thereby avoiding frequent and expensive movement of data between the compute unit and the storage memory. In-memory computing with respect to Silicon memories has been widely explored on various memory bit-cells. Embedding computation inside the 6 transistor (6T) SRAM array is of special interest since it is the most widely used on-chip memory. In this paper, we present a novel in-memory multiplication followed by accumulation operation capable of performing parallel dot products within 6T SRAM without any changes to the standard bitcell. We, further, study the effect of circuit non-idealities and process variations on the accuracy of the LeNet-5 and VGG neural network architectures against the MNIST and CIFAR-10 datasets, respectively. The proposed in-memory dot-product mechanism achieves 88.8\% and 99\% accuracy for the CIFAR-10 and MNIST, respectively. Compared to the standard von Neumann system, the proposed system is 6.24$\times$ better in energy consumption and 9.42$\times$  better in delay. 

\end{abstract}

\begin{IEEEkeywords}
In-memory computing, 6T SRAM Cell, Multiplication, Accumulation, Neural Networks
\end{IEEEkeywords}

\section{Introduction}

For decades the miniaturization of Silicon field effect transistor has been the major driving factor leading to ever increasing on-chip compute capabilities. However, the classical transistor scaling has slowed down as device dimensions are approaching their physical limits. Unfortunately, this imminent end of transistor scaling comes at a time when massive data compute requirements demanded by machine learning and neural network applications have become more important than ever. With this backdrop, a novel paradigm $-$ \textit{in-memory computing}, is being actively investigated by the research community. In-memory computing attempts to embed certain aspects of computations within the memory array. This allows to bypass the well-known memory-wall bottleneck that limits both the throughput and energy-efficiency of state-of-the-art processors \cite{gokhale1995processing},\cite{elliott1992computational}. 
In-memory computing is being actively explored in both Silicon and beyond-Silicon emerging technologies.

Out of various memory technologies SRAM based in-memory computing is of particular interest. SRAMs occupy dominant chip area in state-of-the-art processors. Further, SRAMs are highly sub-banked, therefore, enabling in-memory  fine-grained compute operations within each sub-bank allows massive parallelism both due to high internal memory bandwidth and the existence of multiple sub-banks. Various previous works have investigated different forms of in-memory enabled SRAM banks. These can be broadly classified in two categories $-$ those that enable bit-wise Boolean computations and those that focus on analog-mixed-signal computations within the SRAM array. Bit-wise Boolean computations by enabling two word-lines simultaneously and using modified sensing circuit have been presented in \cite{agrawal2018x},\cite{aga2017compute},\cite{kang2014energy}. Additionally, such bit-wise computation can be complemented with additional digital circuits at the periphery to implement more complex operations like addition  \cite{agrawal2018x}, \cite{dong20170}. On the other hand, analog charge-based compute techniques have been utilized in works like \cite{kang2014energy},\cite{zhang2017memory},\cite{gonugondla2018variation}. Additionally, current based analog convolution for multi-bit dot product computations has been presented in \cite{aga2017compute},\cite{eckert2018neural}. Note, given the complex nature of analog operations, many analog compute proposals in SRAM arrays have either relied on using explicit analog multipliers at the periphery as in \cite{gonugondla2018variation},\cite{shanbhag12018} for the 6T cells or use less-area efficient 8T or 10T cells for binary and multi-bit dot products. In this paper, we present multi-bit multiplication in 6T SRAM cells through charge sharing principle followed by an analog accumulation operation. We further perform an extensive variation analysis to study the approximations induced in the resulting computations due to the analog nature of processing. Subsequently, based on the obtained approximation we implement a deep neural network and analyze the effect of circuit level approximation on the recognition accuracy for CIFAR10 dataset. The key highlights of this work are as follows:

\begin{enumerate}
\item In-SRAM Analog Multiplication: We make use of the SRAM precharge circuit to perform multi-bit analog multiplication by encoding the bit significance in the pulse width of pre-charge pulse.
\item Analog Accumulation: We propose an almost-linear analog accumulator followed by the multiplication operation to further reduce the overall energy consumption as well as increase computing throughput. 
\item Variation Analysis: To ascertain the proposed in-memory computing primitive robustness, we perform extensive variation analysis of the proposed circuitry considering random variations of the transistor threshold voltage. Additionally, we develop a framework to study the effects of such variations on the application task accuracy.
\item System Evaluation and Comparison: We analyze the system level improvements of the proposed work against the baseline (a typical von-Neumann architecture) and the state-of-the-art in-memory computing works in SRAM. 
\end{enumerate}
The rest of the work is organized as follows. 
In Section \ref{sec:Circuit} we elaborate the proposed in-memory multiplication. Section \ref{sec:acc} explains the accumulation scheme.
Circuit simulation results are presented in Section \ref{sec:Circuit_Result}.
In Section \ref{sec:Variation} we analyze the effect of variations on the end application performance.
System level comparisons are presented in Section \ref{sec:System}.
Section \ref{sec:Conclucion} concludes the manuscript.

\section{\textcolor{black}{6T SRAM based Multi-bit Multiplication}}\label{sec:Circuit}
\begin{figure}[t]
\centerline{\includegraphics[width=0.8\columnwidth]{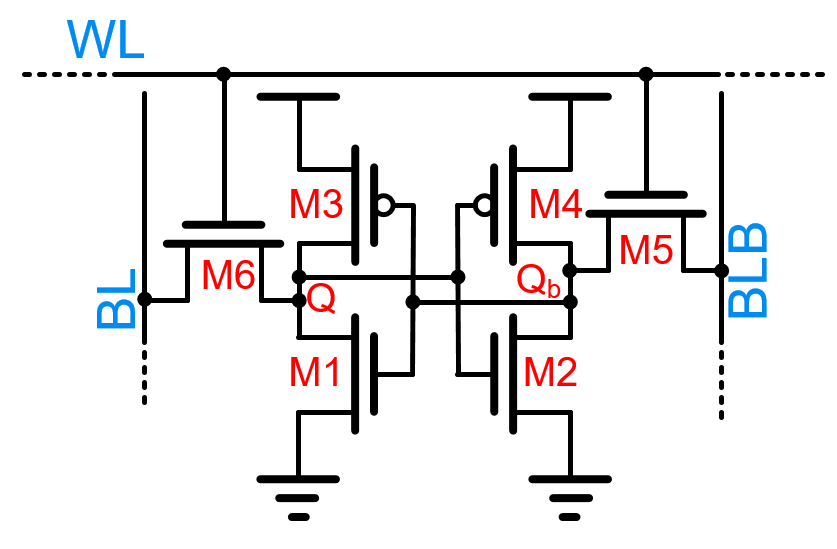}}
\caption{6T-SRAM cell.}
\label{fig:6Tcell}
\end{figure}

We use the standard 6T SRAM cell as the basic memory unit as shown in Fig. \ref{fig:6Tcell}.
The conventional read and write operations in a 6T SRAM cell are performed as follows. For reading the data stored in the cell, the bitline terminals, BL and BLB, are precharged to $V_{DD}$, and the wordline (WL) is enabled. 
When \enquote{1} (Q=$V_{DD}$ and Qb=\unit[0]{V}) is stored in the cell BL voltage remains close to $V_{DD}$, while BLB starts discharging. 
Likewise, when \enquote{0} (Q=0V and Qb=$V{DD}$) is stored BLB would remain close to $V_{DD}$ and BL would discharge from its initial pre-charged voltage. 
For writing \enquote{1} into the cell, BL is pulled to $V_{DD}$, BLB is pulled to 0V and WL is enabled. Similarly for writing \enquote{0} BLB is pulled to $V_{DD}$, BL is pulled to 0V and WL is enabled.

We would now describe the proposed scheme to enable \textit{`in-memory compute'} mode wherein a read operation is accomplished such that instead of reading the individual data, a resultant product between two multi-bit words can be achieved. Specifically, during the multiplication operation one of the operand is passed as input (a voltage on WL) and other operand is stored in memory. 
We denote the operand passed as an input as $V_{in}$ and the operand stored in the memory as W in the rest of the paper.  

\subsection{\textcolor{black}{Multi-bit Single-bit Multiplication in 6T SRAM}}\label{subsection:Vin4W1}
Let us first consider the simpler case, wherein $V_{in}$ is a multi-bit word and W is a single bit word. Suppose W is \enquote{1}(Q=$V_{DD}$ and Qb=\unit[0]{V}) and stored in the 6T SRAM cell. 
As mentioned, BL and BLB are precharged to $V_{DD}$ for a read operation. 
When the world-line is enabled, transistor M5 start conducting but transistor M6 remains in cut-off. 
Thereby transistor M5 creates a path for BLB (pre-charged to $V_{DD}$) to discharge toward ground (0V). 
In fact, the rate of discharge of BLB in this case depends on the discharging current through transistor M5.
The current flowing through the transistor M5 is proportional to its overdrive voltage.
Therefore, changing WL pulse amplitude directly controls the gate voltage of transistor M5, thereby, controlling the discharge rate of BLB. As such, if a multi-bit input is mapped as an analog voltage ranging from the transistor threshold voltage to $V_{DD}$, the discharge on BLB would be proportional to $V_{in}$ provided that Qb is storing 0V. The resultant discharge voltage on the BLB, within a specified time, effectively represents a one bit multiplication on the data stored in SRAM cell (W) with the multi-bit $V_{in}$ applied as an analog voltage on the word-line.
\begin{figure}[b]
\centerline{\includegraphics[width=0.7\columnwidth]{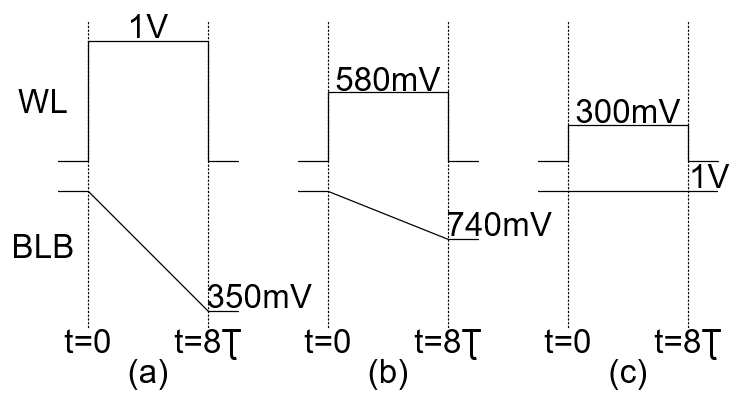}}
\caption{Waveform of WL and BLB voltage when (a) $V_{in}$=15(1111), W=1 (b) $V_{in}$=6(0110), W=1 (c) $V_{in}$=0(0000), W=1.}
\label{fig:4bitVin1bitW}
\end{figure}
\begin{figure*}[t]
  \centering
  \includegraphics[width=0.95\textwidth]{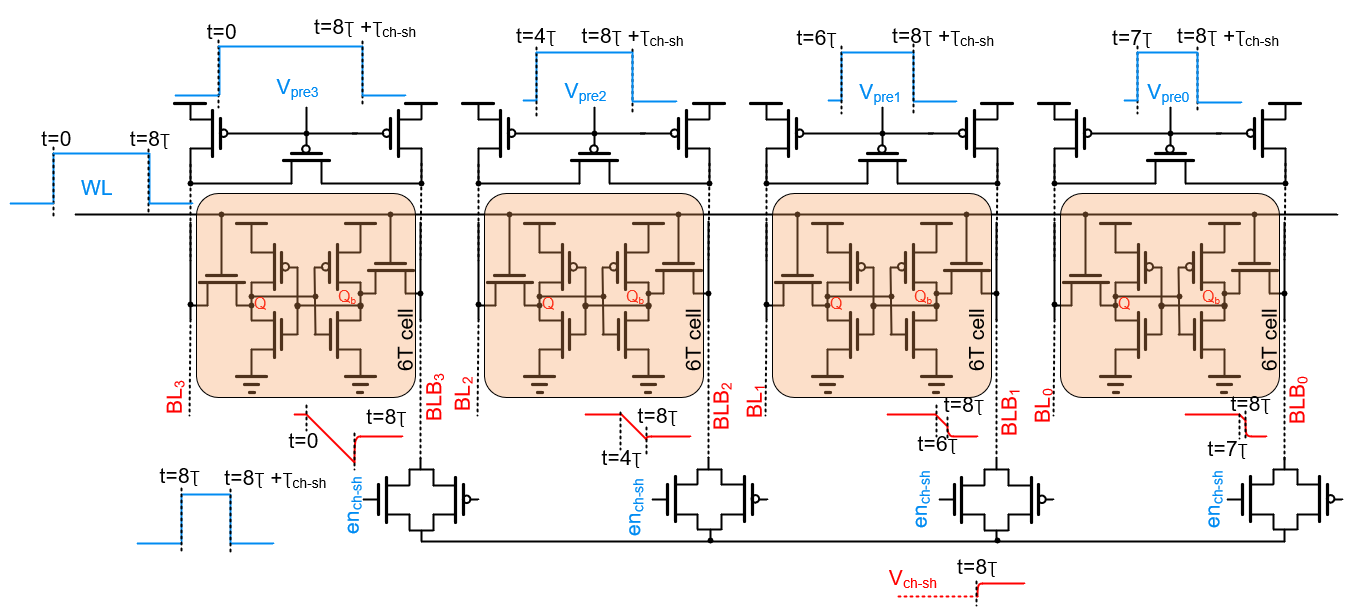}
  \caption{ \textcolor{black}{In-Memory multiplication operation. One of operand W, is stored in the memory cell as shown in the figure. Other operand is encoded as analog voltage on wordline. To capture the bit significance precharge circuits are enabled at different instance of time as shown. The waveforms for WL,} \textcolor{black}{$V_{pre3}$, $V_{pre2}$, $V_{pre1}$, $V_{pre0}$} \textcolor{black}{, $EN_{ch-sh}$, BLB\textsubscript{3}, BLB\textsubscript{2}, BLB\textsubscript{1}, BLB\textsubscript{0}, and $V_{ch-sh}$. $8\uptau$ and $\uptau_{ch-sh}$ are the pulse width of wordline and $EN_{ch-sh}$ signals respectively. Fig. \ref{fig:timingDiagram} shows the waveforms in details.} }
  \label{fig:multiplication}
\end{figure*}

We use a standard SRAM cell in \textcolor{black}{\unit[65]{nm}} TSMC process with BLB discharging rate slowing down drastically at BLB value of $\sim$\textcolor{black}{\unit[100]{mV}}. Thus, we allow BLB to discharge only till \textcolor{black}{\unit[350]{mV}} to maintain the linearity. We refer to the time taken for BLB to discharge from $V_{DD}$ to \textcolor{black}{\unit[350]{mV}} when WL is at $V_{DD}$ to be $8\uptau$.
We limit the maximum WL pulse width to $8\uptau$. 
Assuming a 4-bit word for $V_{in}$, this corresponds to the case when $V_{in}$=15 ($V_{in}$=1111, maximum value) $-$ the amplitude of WL is $V_{DD}$ and pulse-width of WL is $8\uptau$. Fig. \ref{fig:4bitVin1bitW} (a) shows BLB discharge process versus time.
On the other hand, when $V_{in}$=0 ($V_{in}$=0000, minimum value), the amplitude of WL is mapped to \textcolor{black}{\unit[300]{mV}} (close to the value of transistor threshold voltage) while the pulse-width remains $8\uptau$. 
In this case the transistor M5 will no be able to switch ON and BLB will not discharge as shown in Fig. \ref{fig:4bitVin1bitW} (c).
For $V_{in}$ value lying between 0 to 15, say 6, the amplitude of WL lies between the two extreme cases, keeping the pulse width of WL as $8\uptau$.
Here, the transistor M5 switches ON but as the overdrive voltage is low the discharge rate of BLB will be proportionally low. 
As the duration of discharge is fixed to be $8\uptau$, the BLB would not discharge completely and rather settle to an intermediate value as shown in Fig. \ref{fig:4bitVin1bitW} (b). 
Essentially, the discharge rate is proportional to $V_{in}$ and since the duration of discharge is fixed, it can be shown that discharge of BLB after a period of $8\uptau$ is proportional to $V_{in}$.
Conversely, when W is \enquote{0} (Q=0V and Qb=$V_{DD}$) and stored in a 6T SRAM cell as shown in Fig. \ref{fig:6Tcell}, transistor M5 remains in cut off and transistor M6 switches ON.
In this case the source voltage of the M5 is $V_{DD}$ and the gate voltage of the M5 is WL. Thus, at any $V_{in}$, M5 will still be in cut off.

To summarize, when W is \enquote{0} BLB does not discharge and remains close to it's pre-charged voltage.
When W is \enquote{1} BLB discharges by an amount proportional to the input operand $V_{in}$. 
The reason for this being the rate of discharge of BLB is made proportional to $V_{in}$ and the duration of discharge is fixed to $8\uptau$. 
If the analog voltage obtained after converting $V_{in}$ is \textit{v} + \textcolor{black}{\unit[300]{mV}} (since we map the lowest $V_{in}$ to \textcolor{black}{\unit[300]{mV}}) and W(2\textsuperscript{nd} operand) stored in memory is \textit{w}, then the discharge of BLB can be given by following proportionality in equation (\ref{eq:Vin4W1}).

\begin{equation}\label{eq:Vin4W1}
  \begin{aligned}
        V_\text{{precharge}}-{V_\text{BLB}} &\propto \textit{v}\times\textit{w}
  \end{aligned}
\end{equation}

\subsection{\textcolor{black}{Multi-bit Multi-bit Multiplication in 6T SRAM}}\label{subsection:Vin4W4}

In this Section we discuss how to perform multiplication of 4-bit Vin by 4-bit W.
The 4-bit W is stored within 4 adjacent 6T SRAM cells in a row as shown in Fig. \ref{fig:multiplication} with MSB stored in the rightmost cell and LSB stored in the leftmost cell.
In order to achieve multi-bit multiplication with respect to stored value W, the BLB discharge rate should dependent on bit significance, i.e. MSB bit leads to higher discharge as compared to LSB bit. Inducing such bit significance based discharge in conjunction to the discharge rate being proportional to $V_{in}$ enables multi-bit multiplication between Vin and W.

This requirement can be fulfilled through the pre-charge circuits that hold the BL and BLB to $V_{DD}$. We can disable the pre-charge circuit by making $V_{pre}$ voltage equal to $V_{DD}$ at different instant of time such that the BLB corresponding to the MSB bit starts discharging before the LSB bit.
The ratio of time intervals for which MSB, 2\textsuperscript{nd} bit, 3\textsuperscript{rd} bit and LSB should discharge is 8:4:2:1. Thereby, by appropriately pulsing the pre-charge circuit, while $V_{in}$ is applied on the WL, would make the discharge on the BLBs proportional to their bit significance and to the input voltage $V_{in}$.
Interestingly, by enabling the $EN_{ch-sh}$ signal after the BLBs have been given sufficient time to discharge, the four BLBs get connected to each other leading to charge sharing. 
This in turn ensures that the average discharge obtained from 4 BLBs (BLB\textsubscript{3}, BLB\textsubscript{3}, BLB\textsubscript{1}, and BLB\textsubscript{0}) is proportional to analog value of W as explained by equation \ref{eq:Vin4W4}. In the proportionality equation (\ref{eq:Vin4W4}), \textit{w} ($2^3\textit{w}_3+2^2\textit{w}_2+2^1\textit{w}_1+2^0\textit{w}_0$) is the analog equivalent of W. 

\begin{equation}\label{eq:Vin4W4}
  \begin{aligned}
        V_{\text{precharge}}-V_{\text{ch-sh}} &\propto\textit{v}\uptau[2^3\textit{w}_3+2^2\textit{w}_2+2^1\textit{w}_1+2^0\textit{w}_0]/4     \\
               &\propto \textit{v}\times\textit{w}
  \end{aligned}
\end{equation}
\begin{figure}[t]
\centerline{\includegraphics[width=1\columnwidth]{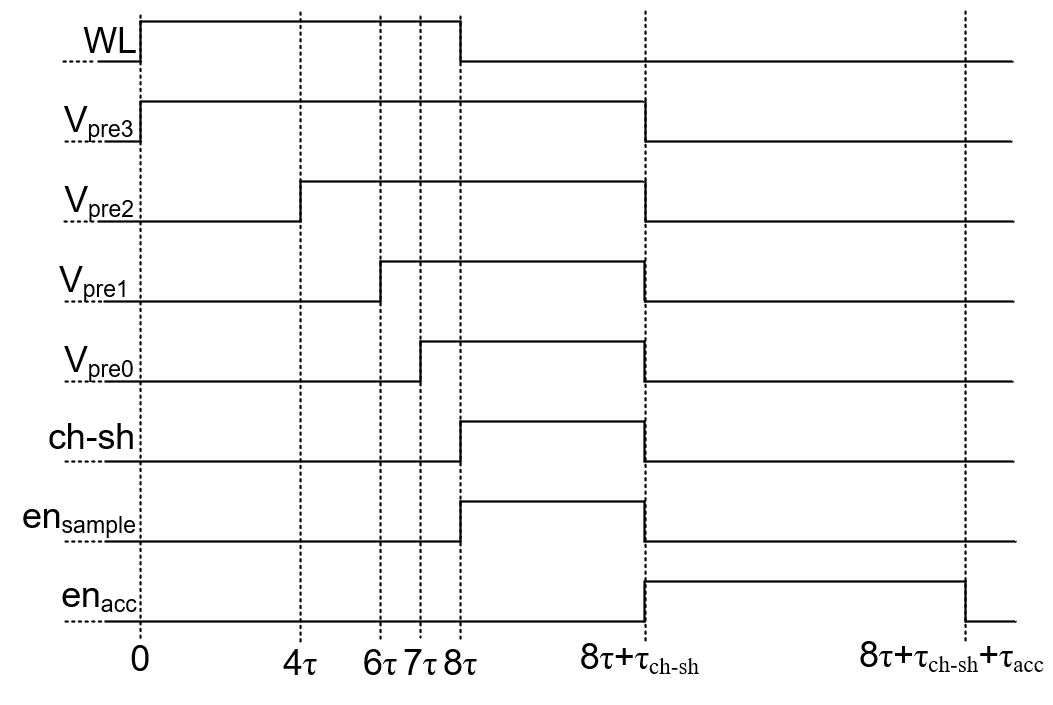}}
\caption{\textcolor{black}{Timing diagram for multiplication operation showing all the signals necessary to perform analog multiplication in \enquote{functional read mode}}}
\label{fig:timingDiagram}
\end{figure}
As mentioned in Section \ref{subsection:Vin4W1}, the pulse width of WL is $8\uptau$ and the amplitude of WL is dependent on $V_{in}$, hence the intervals of discharge for BLB\textsubscript{3}, BLB\textsubscript{3}, BLB\textsubscript{1} and BLB\textsubscript{0} should be $8\uptau$, $4\uptau$, $2\uptau$ and $1\uptau$ respectively.
If WL and \textcolor{black}{$V_{pre3}$} signals are enabled at the instant say t=0, then \textcolor{black}{$V_{pre2}$, $V_{pre1}$ and $V_{pre0}$} are enabled at $4\uptau$, $6\uptau$ and $7\uptau$, respectively.
At t=$8\uptau$, $EN_{ch-sh}$ signal is enabled and has a pulse width of $\uptau_{ch-sh}$.
This enables charge sharing among BLB\textsubscript{3}, BLB\textsubscript{3}, BLB\textsubscript{1} and BLB\textsubscript{0}. 
The voltage obtained after the charge sharing at the node $V_{ch-sh}$ is the analog multiplication result of 4-bit $V_{in}$ and 4-bit W. Timing diagram for various waveforms is shown in Fig. \ref{fig:timingDiagram}. 
\textcolor{black}{In Fig. \ref{fig:timingDiagram} we can see that there is a overlap between the WL signal and the precharge signal($V_{pre3}$). This means that there will be a static current flowing from the PMOS of the precharge through the access transistor. Next, we explain why static current does not play a significant role in the overall energy of the multiplication operation.}
\subsection{\textcolor{black}{Static Current}}\label{subsection:Static_current}
\textcolor{black}{Multiplication operation in our approach is based on constant current discharge of the bit line as explained in Section \ref{subsection:Vin4W1} (Fig. \ref{fig:4bitVin1bitW}). 
The bit line corresponding to the end of the latch storing \enquote{0} will start discharging the bit line through the access transistor, while the other access transistor will remain switched off. In Section \ref{subsection:read_stability} we show that the internal node voltage of the SRAM cell is held close to the initial stored node voltage during the multiplication operation. 
Hence, the discharge current is mainly governed by the access transistor as the node storing \enquote{0} is strongly held close to 0V by the NMOS transistor of the inverter latch. 
Further, to have a nearly constant current discharge we ensure that $V_{DS}$ of the access transistors is large enough to operate in the constant current region of $I_{DS}$ - $V_{DS}$ characteristics. Therefore, we do not allow the bit line voltage to fall below a minimum voltage ($\approx \unit[350]{mV}$), as mentioned in Section \ref{subsection:Vin4W1}, to have a high enough $V_{DS}$ across the access transistor. Regarding the case where the WL signal and the precharge circuit have no overlap, BLB will discharge through the access transistors with a constant current. In this case, the charge stored on the bit line capacitance acts as source of power to facilitate constant current, this in effect reduces the voltage of capacitance at a constant rate. Due to channel length modulation effect the current through the access transistor not perfectly constant as the BLB voltage drops from \unit[1200]{mV} to \unit[350]{mV}, hence, to tackle this problem we change the pulse width of the precharge circuit to make the discharge at BLB3, BLB2, BLB1, and BLB0  in the ration 8:4:2:1 respectively. This approach is discussed in detail in \ref{sec:linearity}.}
	
\textcolor{black}{ The static current discharge path has a PMOS transistor of precharge circuit, NMOS access transistor and the NMOS transistor of the inverter latch. The PMOS transistor of the precharge circuit is strong enough to hold the bit line voltage close to 1.2V, when both the precharge and WL signals are enabled. Hence, the access transistor connected to the internal node storing logic low, has one of the end held to close to ground by the NMOS transistor of the inverter latch circuit, while the other end held close to 1.2V by the PMOS transistor of the precharge circuit. As a result, the current following through the access transistor and the NMOS transistor of the inverter latch is still governed by the access transistors. This current is equal to the fixed current that was discussed in the case where there was no overlap between the precharge circuit operation and the WL signal. This is due to the fact that the current through the access transistor is constant for $V_{DS}$ ranging from \unit[350]{mV} to \unit[1200]{mV} and in this case $V_{DS}$ is $\approx\unit[1200]{mV}$. The difference in this case is that the bit line capacitor maintains it's charge and the constant current in this case is supplied by the power supply through the precharge circuit. }

\textcolor{black}{From the above two cases, it is clear that overlap between the operation of precharge circuit and the WL signal does not play a key role in terms of the energy consumption. Both the case dump equal amounts of charge to the ground if we neglect the channel length modulation effect of the access transistor. The difference in both the cases is that the case having the overlap in the operation of precharge and WL signal takes the charge form the power supply directly, while on the other hand the case where there is no overlap between the two operations takes the charge form the bit line capacitance which in turn is replenished by the supply during the next cycle of operation of precharge circuit. Our circuit simulations in spectre show constant and similar currents values for the $I_{DS}$ of the NMOS  transistor of the inverter latch for both the aforementioned case and justify the above explanation. In the next Section, we describe the array peripherals required to perform Multiply and Accumulate (MAC) operation.  }
    
\section{\textcolor{black}{Peripheral Circuits}}\label{sec:acc}
\textcolor{black}{In this Section we explain the analog accumulator and the SAR-ADC in detail. Further, we provide provide the overall array architecture.}
\subsection{\textcolor{black}{Analog Accumulator}}\label{subsection:Array_peripherals}
\begin{figure}[b]
\centerline{\includegraphics[width=0.7\columnwidth]{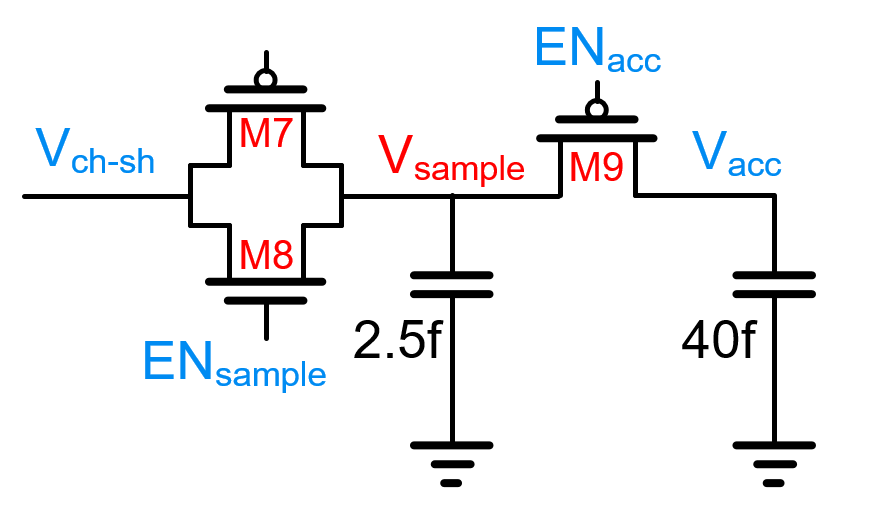}}
\caption{Analog Accumulator used to reduce ADC operations.}
\label{fig:acc}
\end{figure}
\textcolor{black}{The accumulation circuit shown in Fig. \ref{fig:acc} consist of a sample and hold circuit formed by a transmission gate (M7 and M8) and  a sampling capacitor; a PMOS transistor (M9) and an accumulation capacitor. It is important to note that we use PMOS transistor instead of transmission gate for accumulating charge on the accumulation capacitor. Transistor M9 can only conduct when its overdrive voltage is positive, i.e. if 0V is applied at the gate of this transistor, it will only conduct when $V_{sample}$ is greater than the threshold voltage of transistor M9($V_{th,M9}$).	If we are able to maintain the voltage of the accumulation capacitor less than $V_{th,M9}$ for the entire process of accumulation period, then we can see that charge dumped on the accumulation capacitor would be $C_{sample}\times(V_{sample} - V_{th,M9})$. This charge is independent of the capacitor ratios or the initial charge/voltage on the accumulation capacitor. The increase in the voltage of accumulation capacitor would be given by:}
	\begin{equation}
	\label{eqn:deltaV}
	    \color{black}\Delta V_{acc} = C_{sample}\times\frac{(V_{sample} - V_{th,M9})}{C_{acc}}
	\end{equation}
\textcolor{black} {Here, $C_{acc}$ and $C_{sample}$ represent the accumulation capacitor($\unit[40]{fF}$) and the sampling capacitor($\unit[2.5]{fF}$) respectively.
Equation (\ref{eqn:deltaV}) holds true only if the following two conditions are met.}
	\begin{equation}
	\label{eqn:condition1}
	    \color{black}V_{sample} \geq V_{th,M9}
	\end{equation}
	\begin{equation}
	\label{eqn:condition2}
	    \color{black}V_{acc} (n \times \Delta V_{acc}) \leq V_{th,M9}
	\end{equation}
\textcolor{black}{Here, n is the number of accumulations performed. Equations (\ref{eqn:condition1}) and (\ref{eqn:condition2}) are the design constrains which help us decide n, $V_{th,M9}$ and the capacitor sizes used in the accumulation circuit. As analog accumulation unit was introduced to skip the ADC operation for every analog product. A high value of n is hence preferred as it leads to less frequent operation of ADC which in turn helps us save energy and improve the overall latency. On the other side, a high value of number of accumulations leads to a higher value accumulation capacitor evident from the above inequalities (\ref{eqn:condition1} and \ref{eqn:condition2}). A larger accumulation capacitor would occupy more area and would therefore limit the number of accumulations. A high value of number of accumulation also leads to degradation in software accuracy as the error in the analog value accumulated increases with number of accumulations. Further, the time taken for accumulation is 5 times the time taken for multiplication as seen in Table \ref{tab:parameter}. Therefore, for the number of accumulations we explored the multiples of 5 to facilitate the multiplexing of ADC to further reduce the peripheral area if required. Hence, the number of accumulations(10) were chosen to be maximum provided the area and software accuracy are in acceptable limits. We keep the threshold voltage of transistor M9 to be high enough($\approx$\unit[600]{mV}) to allow larger values of n to satisfy the constrain equation (\ref{eqn:condition2}) while not violating constrain equation (\ref{eqn:condition1}). The sampling capacitor $C_{sample}$ was kept to be the low ($\unit[2.5]{fF}$), so that the voltage at the node $V_{sample}$ is almost equal to the voltage at the node $V_{ch-sh}$ and not a scaled version of $V_{ch-sh}$. This means that the range of volatge at node $V_{sample}$ is also \unit[750]{mV} to \unit[1200]{mV}. Using the value of $C_{sample}$, $V_{th.,M9}$ and n in constrain equation (\ref{eqn:condition2}) we get $25 \times 10^{-15} \leq C_{acc}$ when $V_{sample}$ is kept as \unit[1200]{mV} (voltage corresponding to maximum value of $\Delta V_{acc}$, when all  parameter apart from $V_{sample}$ is kept fixed). Hence we chose the accumulation capacitor to be \unit[40]{fF}.}

\subsection{ \textcolor{black}{Successive Approximation Register based ADC}}\label{sec:SAR_ADC}
\begin{figure}[t]
\centerline{\includegraphics[width=0.9\columnwidth]{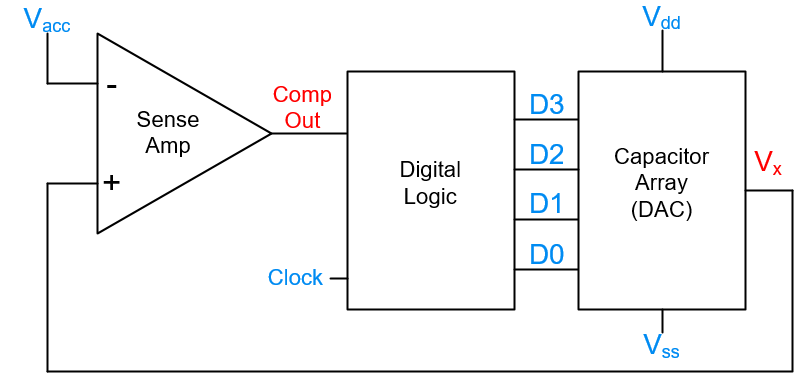}}
\caption{Successive approximation ADC.}
\label{fig:SAR_ADC}
\end{figure}

The accumulated voltage is then converted to a digital output using a Successive Approximation Register ADC (SAR-ADC) as shown in Fig \ref{fig:SAR_ADC}. 
Three components of this ADC are the sense amplifier, digital logic and the capacitor array based Digital to Analog Converter (DAC).
The digital logic initializes the MSB of the ADC output to \enquote{1}, keeping all other bits \enquote{0} and sends the digital code to DAC.
The DAC converts this digital code to an analog voltage ($V_{x}$).
Subsequently, $V_{x}$ is compared against the input analog signal using a comparator.
This is equivalent to comparing the input signal ($V_{acc}$) to $(V_{dd}+V_{ss})/2$.
If the input signal is higher than the DAC output (Compout is 0), MSB is kept at \enquote{1} otherwise, it is flipped to \enquote{0} which concludes 1 cycle of operation for SAR-ADC.
In the consecutive cycle, the digital logic makes the next most significant bit \enquote{1} keeping all the bits following it to be \enquote{0}. 
DAC converts the digital code to the analog signal which is compared against the input signal using the Sense Amplifier.
The sense amplifier output is used to fix the bit under evaluation.
The same process continues until all bits are evaluated.
SAR-ADC takes n-cycles to evaluate the digital output, where n is the bit precision of the ADC.
We adopt a capacitor array as the DAC and a standard sense amplifier as the comparator to reduce the static energy consumption in the utilized SAR-ADC.
The target application of this work inspires the ADC bit-precision to be 4 bits with a conversion delay of \textcolor{black}{\unit[5]{ns}}.

\subsection{Array Overview}\label{subsection:overview}
\begin{figure}[t]
\includegraphics[width=1.0\columnwidth]{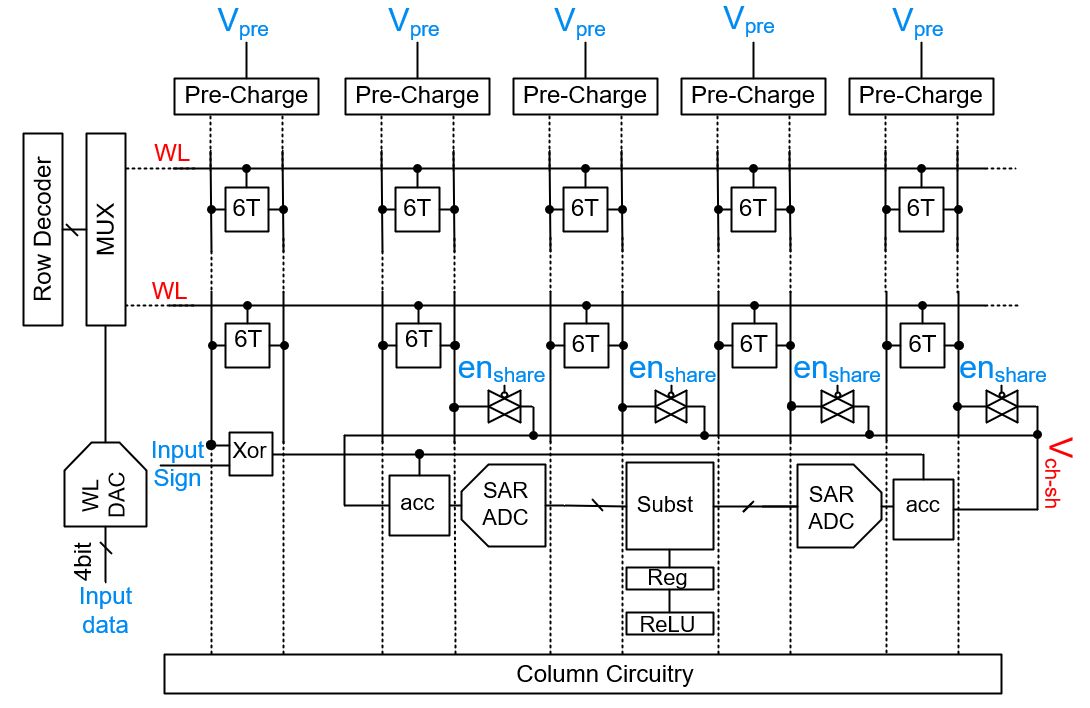}
\caption{6T-SRAM Array overview.}
\label{fig:overall}
\end{figure}

The SRAM array for the proposed in-memory computing methodology is shown in Fig. \ref{fig:overall}. 
We adopt TSMC \unit[65]{nm} technology node in our design. Additionally, we consider \unit[50]{fF} bitline capacitance resulting from metal lines in layout.

The proposed array behaves as a conventional 6T-SRAM memory when $EN_{ch-sh}$ and the analog multiplexer (MUX) are disabled and normal read/write operations are performed.
Moreover, the circuit is capable of performing 4-bit$\times$4-bit multiplication and accumulation to provide a 4-bit output. 
The sign of the analog product is computed as XOR of the signs of both $V_{in}$ and W, and the product is accumulated on two different accumulation capacitors depending on the output sign. Furthermore, we share the circuit peripherals between two $256\times256$  arrays to reduce the area penalty resulting for peripheral circuitry. 

\section{Circuit Simulation Results}\label{sec:Circuit_Result}
\textcolor{black}{In this Section we present circuit simulation results, wherein we analyse the non-linearity in the proposed multiplication circuitry and propose a mechanism to mitigate the non-linearity.}
\subsection{ \textcolor{black}{Linearity Analysis}}\label{subsection:methodology}
\textcolor{black}{The proposed multiplication circuitry, multiplies two scalars, namely, the input $V_{in}$ and the weight $W$ in an analog fashion. For the purpose of the linearity analysis we ignore the signs of both the input and the weight. This is because our approach computes the sign of the result depending on the signs of input and weight using digital gates and is free from any analog non-linearity. On the x-axis in Fig. \ref{fig:nonlinear} we have the  \enquote{Expected Product} or expected multiplication result for the digital multiplication of input and weight, while on y-axis we have the \enquote{Observed Product at $V_{ch-sh}$ (V)} which is the analog product obtained at the node $V_{ch-sh}$ corresponding to input $V_{in}$ and the weight $W$. For example, if the input $V_{in}$ is 5 and the weight $W$ is 10, then the x-coordinate for such a pair will be 50 (=5*10), while the y-coordinate will be the analog multiplication result obtained at $V_{ch-sh}$ by our proposed approach. We use spectre simulation to obtain the analog multiplication values as proposed in the paper. The SRAM bit cells are initialized to the weight value $W$ and input $V_{in}$ is provided to the input DAC. 
The DAC converts the input to an anlog value according to the equation \ref{eq:linear_dac}. To multiply the two scalars $V_{in}$ and $W$, proper signals were applied to precharge circuit ($V_{pre}3$, $V_{pre}2$, $V_{pre}1$ and $V_{pre}0$) and the charge sharing circuit ($En_{ch-sh}$), as shown in Fig. 4. To multiply the two scalars $V_{in}$ and $W$, proper signals were applied to precharge circuit ($V_{pre}3$, $V_{pre}2$, $V_{pre}1$ and $V_{pre}0$) and the charge sharing circuit ($En_{ch-sh}$), as shown in Fig. 4. }
\begin{equation}\label{eq:linear_dac}
  \begin{aligned}
      \color{black}V_{WL}=\unit[300 + V_{in}\times\frac{700}{15}]{mV}
  \end{aligned}
\end{equation}
\textcolor{black}{For the plot shown in the Fig. \ref{fig:nonlinear}, we keep the precharge pulse widths corresponding to $V_{pre3}$, $V_{pre2}$, $V_{pre1}$, and $V_{pre0}$ such that the duration for which the BLB discharges are in the ratio 8:4:2:1 corresponding to BLB3, BLB2, BLB1 and BLB0 respectively (as shown in Fig. 4). However, with this approach we see from the scatter plot that the analog product obtained is non-linear and would require some correction techniques to mitigate the error. The possible options could be to have a non linear mapping for the input $V_{in}$ which would require design of a complicated DAC structure or non linear mapping of the weights which would require storing the non linear weights in the memory. In the next subsection we discuss how we mitigate this non linearity without having to use non-linear mapping.}
{\centering
\begin{figure}[t]
        \centering
         \includegraphics[width=0.65\columnwidth]{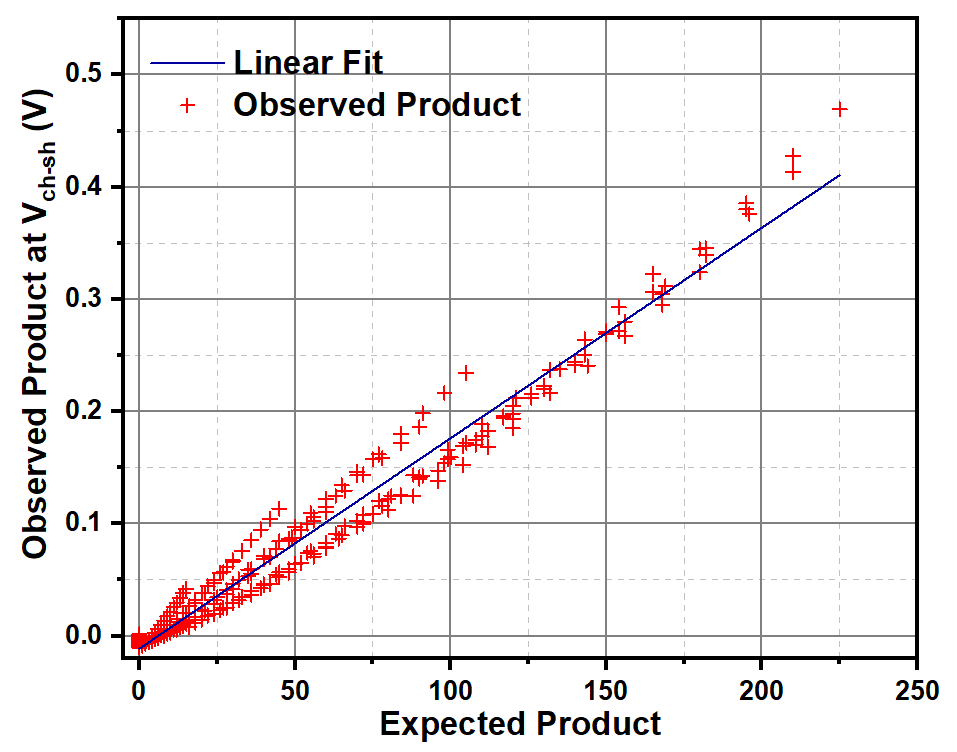}
         \caption{Analog Output when Pre-charge pulse width ratios 8:4:2:1.}
         \label{fig:nonlinear}
\end{figure}
}
{\centering
\begin{figure}[b]
        \centering
         \includegraphics[width=0.65\columnwidth]{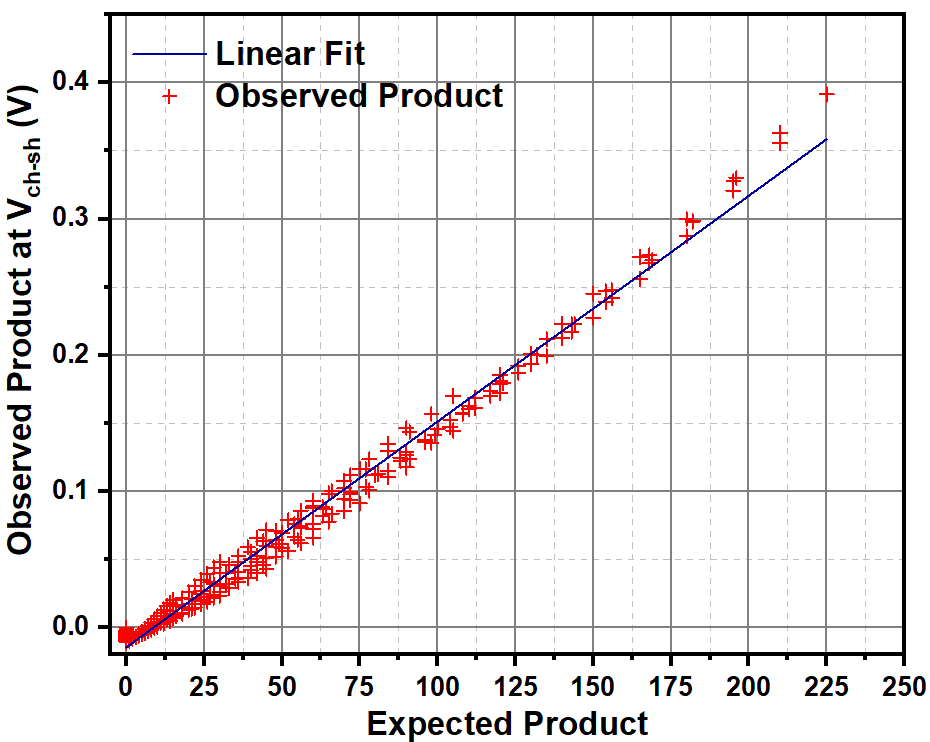}
         \caption{Analog Output when BLB discharge ratio 8:4:2:1.}
         \label{fig:linear}
\end{figure}
}
\subsection{\textcolor{black}{Linearity Enhancement}}\label{sec:linearity}
\textcolor{black}{We saw that the plot in Fig. \ref{fig:nonlinear} could be explained with four lines. This systematic non-linearity caused us to examine the analog multiplication even further. The approach followed to obtain results as shown in Fig. \ref{fig:nonlinear} is based on the assumption that the discharge of the BLB is the proportional to duration of discharge. However, we know that the access transistors in the SRAM bit cell are non-linear and hence the above assumption would no longer hold true. For equation (\ref{eq:Vin4W4}) to hold, the discharge of BLB3, BLB2, BLB1 and BLB0 should be in ratio 8:4:2:1. Therefore, we keep the duration for which BLB3, BLB2, BLB1 and BLB0 discharge such that the discharge on the corresponding BLB is in the ratio 8:4:2:1 respectively. In other words, rather than keeping the duration of discharge in ratio 8:4:2:1, the discharge is kept in the ratio 8:4:2:1. To obtain the pulse width of $V_{pre3}$, $V_{pre2}$, $V_{pre1}$, and $_{pre0}$, we obtain the time taken by the BLB to discharge to \unit[350]{mV} (minimum voltage for almost linear discharge as explained in Section \ref{sec:Circuit}), \unit[775]{mV}, \unit[987.5]{mV} and \unit[1093.75]{mV} respectively from 1.2V when the bit stored in bit cell is high which makes the discharge in the ratio 8(1200-350):4(1200-775):2(1200-987.5):1(1200-1093.75) respectively. 
For further analysis the ratios of the pre-charge pulse width are kept such that the BLB discharge ratio is 8:4:2:1.}

\begin{figure}[t]
     \centering
     \begin{subfigure}[htbp]{0.65\columnwidth}
         \centering
         \includegraphics[width=\columnwidth]{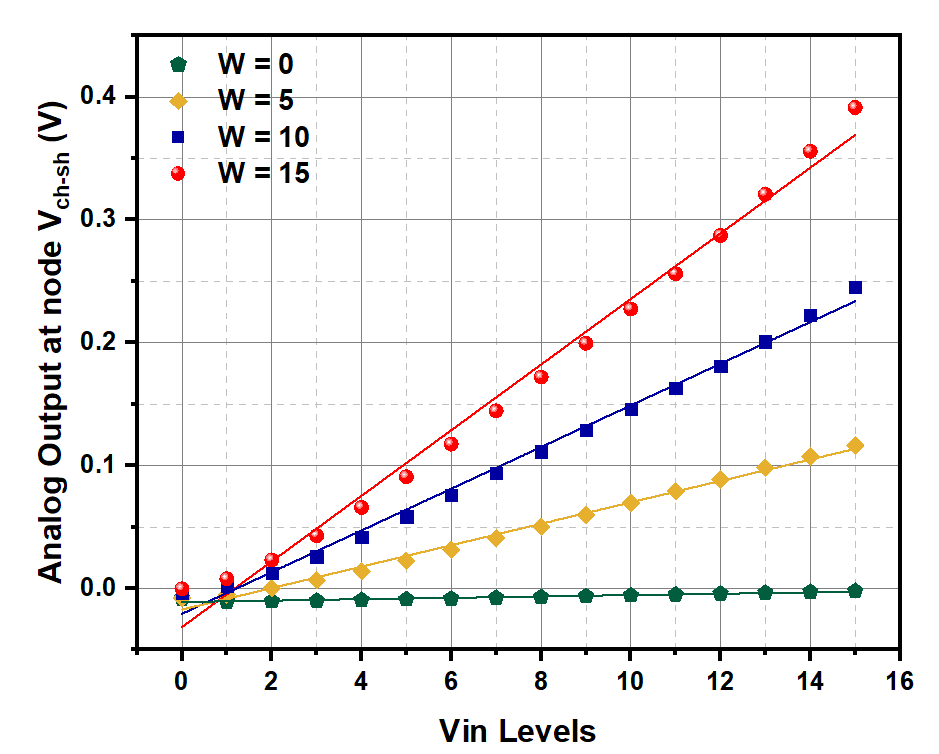}
         \caption{}
         \label{fig:OutVsVin}
     \end{subfigure}
     \begin{subfigure}[htbp]{0.65\columnwidth}
         \centering
         \includegraphics[width=\columnwidth]{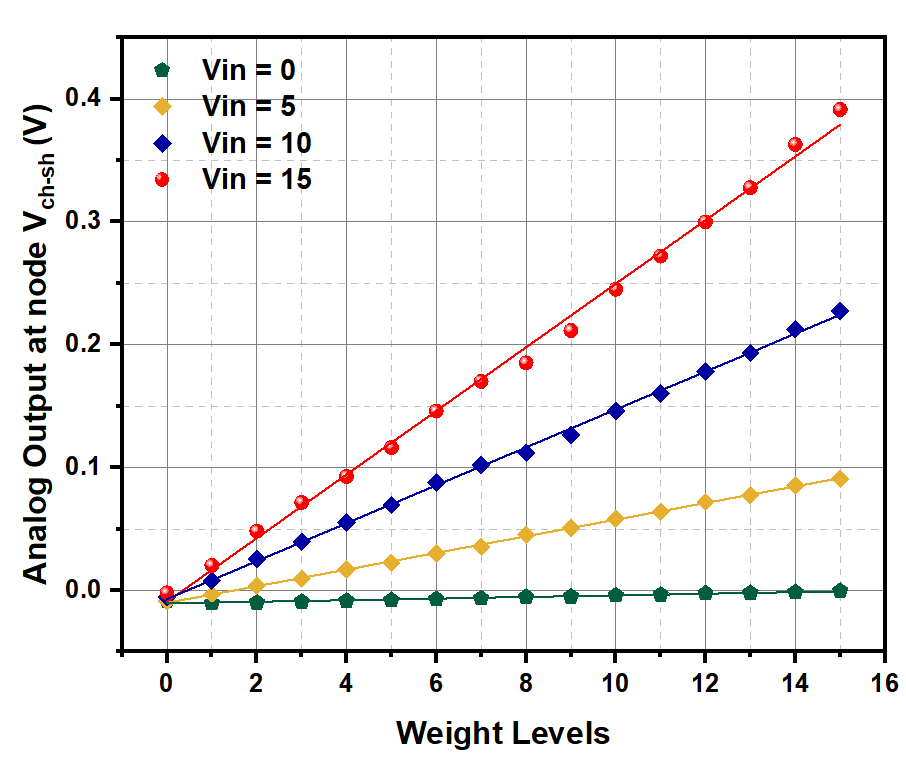}
         \caption{}
         \label{fig:OutVsW}
     \end{subfigure}
        \caption{(a) Effect of input $V_{in}$ on the analog output for various W. (b) Effect of W on the analog output for various $V_{in}$.}
        \label{fig:analog_mult}
\end{figure}

In Fig. \ref{fig:OutVsVin} we present the analog multiplication output with varying $V_{in}$ and W = 0, 5, 10 and 15, similar trends were observed for other values of W.
Similarly, in Fig. \ref{fig:OutVsW} we observe the analog multiplication output with varying W and $V_{in}$ fixed to 0, 5, 10 and 15, similar trend observed for other values of $V_{in}$.  Fig. \ref{fig:OutVsVin} and Fig.  \ref{fig:OutVsW} show that digital output is directly proportional to the input voltage $V_{in}$ and the $W$ stored in the memory, respectively, confirming the multiplication operation of the proposed scheme. \textcolor{black}{For Fig. \ref{fig:OutVsVin} the Integral Non-Linearity(INL) values for the W 0, 5, 10 and 15 are 1.13$\times$10\textsuperscript{-14}, 1.11$\times$10\textsuperscript{-14}, -4.88$\times$10\textsuperscript{-14} and 2.55$\times$10\textsuperscript{-14} respectively and INL values for $V_{in}$ 0, 5, 10 and 15 in Fig.\ref{fig:OutVsW} are 1.09$\times$10\textsuperscript{-14}, 5.77$\times$10\textsuperscript{-14}, -2.36$\times$10\textsuperscript{-14} and 3.109$\times$10\textsuperscript{-14} respectively.}

Note, the accumulation voltage is always less than the threshold voltage of transistor M9 as discussed in section \ref{subsection:Array_peripherals}. As a result, the analog accumulator behaves linearly by design.
It is worth mentioning that the DAC block in ADC is designed to perform linearly during successive approximation by fine tuning the capacitor ratio. Additionally, the comparator is designed with an offset voltage as low as \textcolor{black}{\unit[5]{mV}}, which is lower than the ADC precision, to ensure linearity.
\section{Variation Analysis}\label{sec:Variation}
\textcolor{black}{In this section we analyse the read stability of the SRAM cells to the \enquote{functional read} in Subsection A. We study the effect of variations on the proposed circuitry considering random variations of the transistor threshold voltage in Subsection B and  develop a framework presented in Subsection C to study the effects of such variations on the end application accuracy. }
\vspace{-0.35cm}
\subsection{\textcolor{black}{Read Stability}}\label{subsection:read_stability}
\textcolor{black}{To avoid read disturb we design the cell to have acceptable read noise margin, following a standard approach for SRAM cell design \cite{mukherjee2010static}. In our approach we only activate one WL at a time which further decreases the chances of corrupt read. When there is overlap between the WL signal and operation of the precharge, both BL and BLB are held high by the precharge circuit and hence the differential signal applied to the cell is zero which ensure that the data stored in the cell is not corrupted by the multiplication operation. When the precharge circuit is disabled, while the WL signal is still high, the circuit performs normal conventional read and discharges BL or BLB depending on data stored in bit cell. Many in-memory compute primitives \cite{shanbhag-arch}, \cite{zhang2017memory}, \cite{verma2016} have shown similar kind of functional read operations where they connect the internal storage node to the bit line in order to enable analog in-memory compute.}

\textcolor{black}{In order to understand the read disturb that could arise due to analog multiplication operation as proposed in our approach, we need to understand the differences in our approach and the conventional memory read operation. In our approach, we have an overlap between the operation of precharge circuit and the WL signal for certain bits of weights, as seen in waveforms shown in Fig. 4. The case where the precharge circuit operation and WL signal have no overlap, as in the case of $V_{pre3}$, the operation is similar to conventional memory read operation. On the other hand, when the precharge circuit operation and WL signal have maximum overlap as seen in the case of $V_{pre0}$, is different from the conventional read case and requires further analysis. We also use analog signal on the WL which is different from the conventional read operation. The worst case WL signal amplitude would be \unit[1]{V}, which is the strongest connection between the internal node and the bit line. Hence, we analyse the case where the amplitude of WL signal is \unit[1]{V} and there is maximum overlap between the operation of the precharge circuit and the WL signal.}

\textcolor{black}{To study the read disturb for these conditions, we simulate two SRAM bit cells with the precharge circuit and initialize the two cells to store opposite bits(one storing digital high and other storing digital low). We perform 10000 Monte Carlo runs, on the two SRAM cells, having the WL amplitude 1V and setting the precharge signal to $V_{pre0}$ and monitor the number of times the bit stored in the internal node switches. We find that none of the cells flip there initial stored states. Further, the maximum change in voltage during the functional read operation at node storing digital low (\unit[0]{V}) is \unit[95.56]{mV} and maximum change in the node storing digital high (\unit[1]{V}) is \unit[51.91]{mV}. This analysis shows that the proposed SRAM in memory multiplication approach is robust to read disturbs.}
\subsection{\textcolor{black}{Variations in Multiplication In-memory}}\label{subsection:Variation} 
Random variations of access transistor threshold voltage are one of the major causes of performance degradation in circuit design. In the proposed circuit, the mismatch between BLB access transistors of the 4 adjacent SRAM cells, shown in Fig. \ref{fig:multiplication}, can cause erroneous output.
Such output errors happen because the rate of discharge for BLB\textsubscript{3}, BLB\textsubscript{2}, BLB\textsubscript{1} and BLB\textsubscript{0} are no longer proportional to 8:4:2:1 for the same value of input $V_{in}$. We perform Monte Carlo runs using the TSMC \textcolor{black}{\unit[65]{nm}} PDK to simulate the random variations in the proposed circuitry.
The maximum standard deviations for the analog multiplication output in 1000 Monte Carlo runs is seen to be \textcolor{black}{\unit[13.17]{mV}}.

To assess the effect of variations  on the digital output we run 1000 samples of Monte Carlo simulation on the entire circuit consisting of SRAM array and the peripheral circuits including the ADC. 
The simulation results for a chosen set of cases where the digital output is 0,3,6,8,11 and 15 are shown in Fig. \ref{fig:MC}. 
Notice that the maximum standard deviation of the Gaussian curve for digital output was 0.6.
From Fig. \ref{fig:MC} we see that the distributions for digital output 0 and 15 are truncated Gaussian distributions. The reason for that is the ADC being only able to map the input voltage in its predefined range of conversion.
Any voltage below the range of conversion is mapped to 0 and any voltage above the range of conversion would be mapped to 15.
{\centering
\begin{figure}[t]
\centering\includegraphics[width=1.\columnwidth]{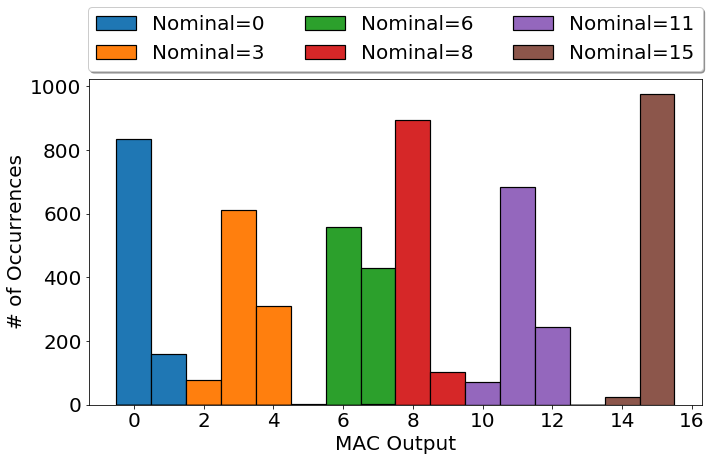}
\caption{Histogram showing the effect of process variation on final digital output after the 10 element 4bit x 4bit multiply and accumulate operation.}
\label{fig:MC}
\end{figure}
}
\subsection{Functional Accuracy Analysis}
We test the proposed multiplication and accumulation engine on a convolutional neural network shown in Table \ref{tab:VGG} for classifying Cifar10 dataset. We refer to \cite{github} for the training framework.
The network is trained using Adam optimizer with initial learning rate of 0.0001 for 200 epochs.
The learning rate is dropped by a factor of 10 at epoch number 100, 150 and 180. 
The training batch size and testing batch size is 32 and 128 respectively.
We perform small data augmentation (Flipping the training dataset and shifting the training dataset by 4 pixels).
Cross-entropy loss function is adopted for training, while the test accuracy for such network is 91.50\%.
Now, in order to assess the effect of hardware non ideal effects we first run the neural network with quantized weights and activation with the weights and activations being quantized separately for each layer of the network.
Using linear quantization, the minimum number of bits required to represent activations and weights is 5-bits to get acceptable accuracy of 90.91\%.
Moreover, we add an effective multiplication error to the output map sampled from a Gaussian distribution of circuit variation sigma equal to 0.6$\times\sqrt{n}$, where n is number of 10 element MAC operations and standard deviation for individual multiplication from the variation analysis was taken to be 0.6 following a pessimistic approach.
For inference, we sample the errors from the Gaussian distribution for each output map and keep it constant for that particular inference as the error is dependent on the location where the weight is written and is fixed once the weight is stored in SRAM. 
We run 1000 inferences adding error to the output map as described above, the test accuracy remains between 88.83\%-89.62\% (\textcolor{black}{mean=89.25\% and sigma=0.1329}) with circuit variations.

{\centering\begin{table}[t]
    \caption{VGG used for classifying Cifar10 dataset} 
    \label{tab:VGG}
    \centering
         \begin{tabular}
          {||c c c c||} 
         \hline
         Layer & Input Map & Output Map & Non Linearity \\ [0.5ex] 
         \hline\hline
         64 3x3 Conv1 & 32x32x3 & 32x32x64 & ReLU,dropout(0.3) \\ 
         \hline
         64 3x3 Conv2 & 32x32x64 & 32x32x64 & ReLU \\
         \hline
         [2 2] MaxPool1 & 32x32x64 & 16x16x64 & - \\
         \hline
         128 3x3 Conv3 & 16x16x64 & 16x16x128 & ReLU,dropout(0.4)  \\
         \hline
         128 3x3 Conv4 & 16x16x128 & 16x16x128 & ReLU  \\
         \hline
         [2 2] MaxPool2 & 16x16x128 & 8x8x128 & - \\
         \hline
         256 3x3 Conv5 & 8x8x128 & 8x8x256 & ReLU,dropout(0.4) \\
         \hline
         256 3x3 Conv6 & 8x8x256 & 8x8x256 & ReLU,dropout(0.4) \\
         \hline
         256 3x3 Conv7 & 8x8x256 & 8x8x256 & ReLU \\  
         \hline
         [2 2] MaxPool3 & 8x8x256 & 4x4x256 & - \\
         \hline
         4096x4096 FC1 & 1x1x4096 & 1x1x4096 & ReLU,dropout(0.5) \\
         \hline
         4096x4096 FC2 & 1x1x4096 & 1x1x4096 & ReLU,dropout(0.5) \\
         \hline
         4096x10 FC3 & 1x1x4096 & 1x1x10 & -\\
         \hline
        \end{tabular}
\end{table}
}

Most of the recent in-memory work in literature are designed to run neural networks against the MNIST dataset.
We use our network model to classify MNIST dataset on a vanilla LeNet network.
For this task full precision test accuracy is found to be 99.3\%, quantized (4-bit activation and 4-bit weights) test accuracy is 99.24\% and test accuracy for our approach remains between 99.05\% to 99.32\% (\textcolor{black}{mean=99.19\% and sigma=0.0398}) for 1000 variation-affected inference runs.

\section{\textcolor{black}{System level Analysis}}\label{sec:System}

{\centering\begin{table*}[t]
    \caption{Comparison with other related works} 
    \label{tab:compare}
    \centering
         \begin{tabular}
          {|| c | c | c | c | c | c | c | c | c | c ||} 
         \hline
         \multicolumn{1}{||p{1cm}|}{\centering Related \\ Work }
         & \multicolumn{1}{|p{1cm}|}{\centering Process \\ (nm) } 
         & \multicolumn{1}{|p{1cm}|}{\centering SRAM\\ Array }
         & \multicolumn{1}{|p{1cm}|}{\centering Cell\\ Area }
         & \multicolumn{1}{|p{1.5cm}|}{\centering Computation\\ Nature}
         & \multicolumn{1}{|p{1.5cm}|}{\centering Input/Weight\\ precision }
         & \multicolumn{1}{|p{1cm}|}{\centering Dataset }
         & \multicolumn{1}{|p{1.5cm}|}{\centering Network\\ Architecture }
         & \multicolumn{1}{|p{1cm}|}{\centering Accuracy } 
         & \multicolumn{1}{|p{3cm}||}{\centering Energy}\\[0.5ex] 
         \hline\hline
         IMAC & 65 & 6T & 1x & Analog & 5/5 & \multicolumn{1}{|p{1cm}|}{\centering Cifar10\\ MNIST } & \multicolumn{1}{|p{1.5cm}|}{\centering VGG \\ LeNet-5} & \multicolumn{1}{|p{1.5cm}|}{\centering >88.83\%\\>99.05\%  } & \multicolumn{1}{|p{3cm}||}{\centering -\\ \textcolor{black}{\unit[158.203]{nJ}} (/inference)}\\ 
         \hline
         \cite{shanbhag-arch}& 65 & 6T & 1x & Analog & 6/4 & MNIST & LeNet-5 & >97\% & \textcolor{black}{\unit[359.288]{nJ}} (/inference)\\ 
         \hline
         \cite{jaiswal20188t} & 45 & 8T & 1.68x & Analog & -/4 & MNIST & MLP & 98.15\% & $\sim$\textcolor{black}{\unit[180]{pJ}} (/16$\times$16 MAC)\\ 
         \hline
         \cite{sandwitch-ram}& 28 & 8T & 1.3x & Analog & 8/1 & ImageNet & AlexNet & <1\% drop &  \textcolor{black}{\unit[12.8-119.7] {TOPS/W}} \\ 
         \hline
         \cite{jiang2018xnor}& 65 & 12T & 2x & Analog & ternary/1 & \multicolumn{1}{|p{1cm}|}{\centering Cifar10\\ MNIST } & \multicolumn{1}{|p{1.5cm}|}{\centering BNN\cite{binarybengio} \\ MLP } & \multicolumn{1}{|p{1.5cm}|}{\centering 85.7\% \\ 98.3\% }& \multicolumn{1}{|p{3cm}||}{\centering  \textcolor{black}{\unit[2.48-7.19]{fJ}}  (/operation)} \\ 
         \hline
         \cite{twin-8t}& 55 & Twin-8T & 1.4x & Digital & 4/5 & \multicolumn{1}{|p{1cm}|}{\centering Cifar10\\MNIST } & \multicolumn{1}{|p{1.5cm}|}{\centering CNN\\CNN } & \multicolumn{1}{|p{1.5cm}|}{\centering 90.42\%\\99.52\%  } & \multicolumn{1}{|p{3cm}||}{\centering  \textcolor{black}{\unit[11.7]{pJ}} (/unit-Macro (64$\times$60 bits))}\\ 
         \hline
         \cite{eckert2018neural}& 28 & 6T & 1x & Digital & 8/8 & \multicolumn{1}{|p{1cm}|}{\centering ImageNet } & \multicolumn{1}{|p{1.5cm}|}{\centering Inception V3 } & \multicolumn{1}{|p{1.5cm}|}{\centering -} & \multicolumn{1}{|p{3cm}||}{\centering  \textcolor{black}{\unit[0.246]{J}} (/inference)}\\ 
         \hline
         \end{tabular}
\end{table*}
}

\begin{figure*}[htbp]
\minipage{0.24\textwidth}
  \includegraphics[width=\linewidth]{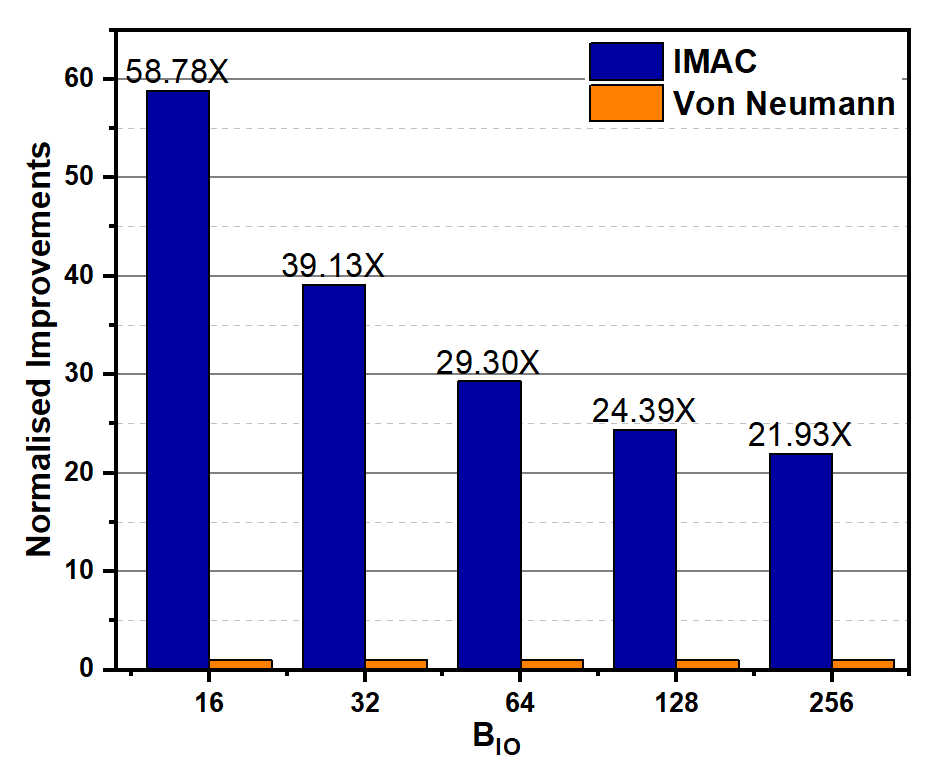}
  \caption{Energy Delay Product comparison for IMAC and von Neumann with varying $B_{IO}$ (bits fetched from SRAM to processor per bank)}\label{fig:EDP_compare}
\endminipage\hfill
\minipage{0.73\textwidth}
     \centering
     \begin{subfigure}[htbp]{0.32\linewidth}
         \centering
         \includegraphics[width=\linewidth]{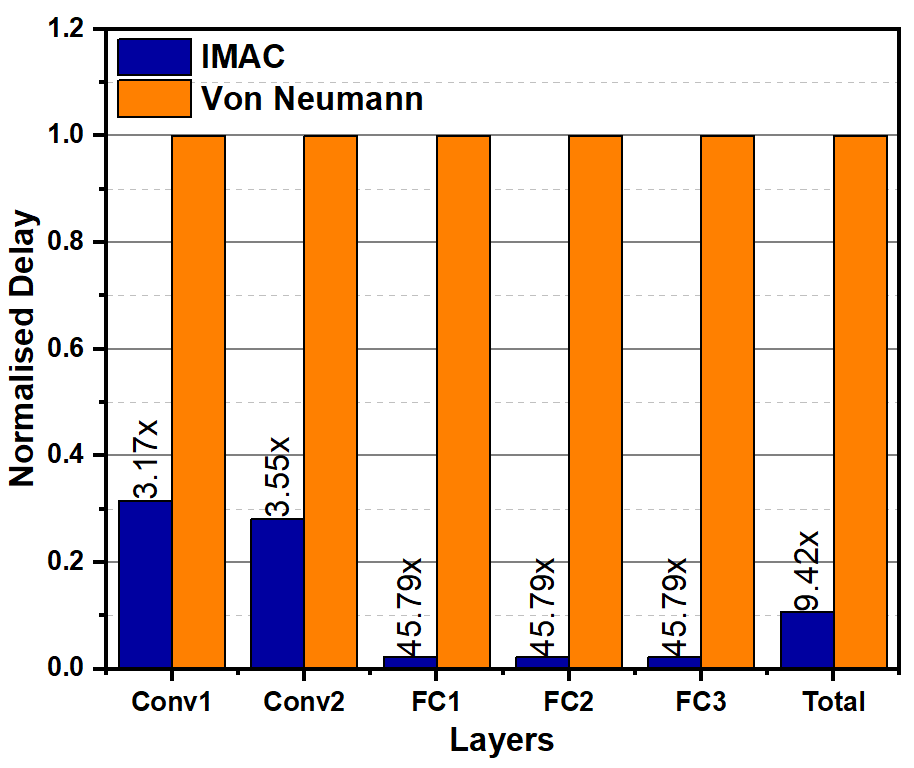}
         \caption{}
         \label{fig:E}
     \end{subfigure}
     \begin{subfigure}[htbp]{0.32\linewidth}
         \centering
         \includegraphics[width=\linewidth]{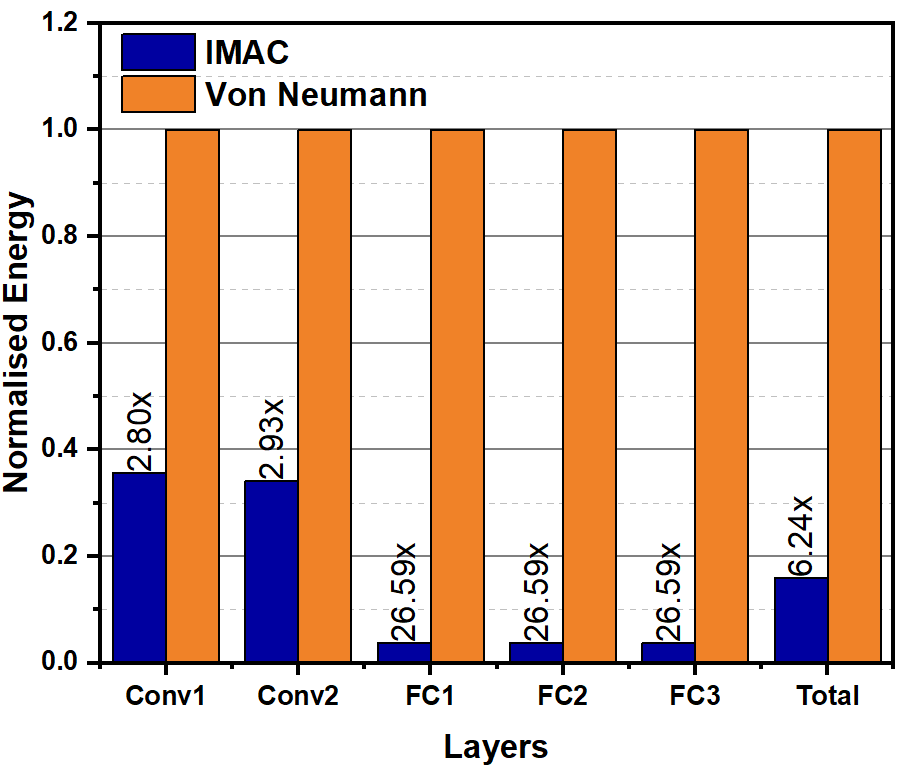}
         \caption{}
         \label{fig:D}
     \end{subfigure}
     \begin{subfigure}[htbp]{0.32\linewidth}
         \centering
         \includegraphics[width=\linewidth]{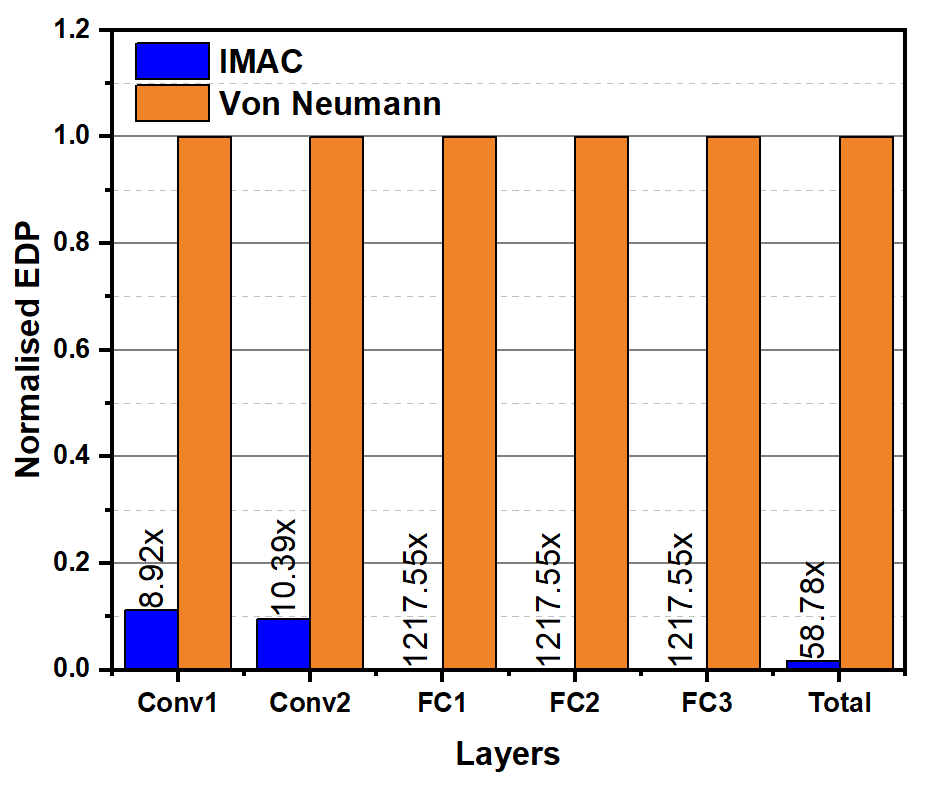}
         \caption{}
         \label{fig:EDP}
     \end{subfigure}
        \caption{ Plots showing inference comparison for different layers in Lenet-5 using MNIST dataset (a)Delay (b) Energy (c) Energy-Delay product.}
        \label{fig:system}
\endminipage
\end{figure*}
In this section we compare the proposed work with a standard von Neumann architecture which is considered as our baseline.
For conventional von Neumann architecture a neural network application can be broken down into Read, Multiply, Accumulate and non-linearity(ReLU in our case) kernels. It is worth mentioning that the ReLU operation energy consumption can be neglected compared to other operations.
The delay ($T_{VN}$) and the energy ($E_{VN}$) of convolution operation in the conventional von Neumann architecture are given by equation (\ref{eq:timeVN}) and (\ref{eq:energyVN}) similar to \cite{shanbhag-arch}.
The notations used in the equations below are described in Table \ref{tab:Notation}.
We can obtain the formula for fully connected layer by simply making the input size(L) and kernel size(K) as 1. In the fully connected layer M and N denote the number of input and output neurons respectively.

\begin{equation}\label{eq:timeVN}
  \begin{aligned}
        T_{VN}\approx \bigg[\frac{MNK^2}{(B_{IO}/B_{W})N_{bank}}\bigg]T_{read}+ \bigg[\frac{MNK^2}{N_{mult}}\bigg]N_{mov}^2T_{mult}
  \end{aligned}
\end{equation}

\begin{equation}\label{eq:energyVN}
  \begin{aligned}
        E_{VN}\approx MNK^2E_{read}
        + MNK^2N_{mov}^2E_{mult}
        + P_{leak}T_{VN}
  \end{aligned}
\end{equation}

The delay($T_{in-memory}$) and energy($E_{in-memory}$) for convolution operation involving in-memory multiplication operation as described in this work is given by equation (\ref{eq:time}) and (\ref{eq:energy}) respectively.

{\centering\begin{table}[t]
    \caption{Notations used in the delay and energy equation} 
    \label{tab:Notation}
    \resizebox{\columnwidth}{!}{
    \centering
         \begin{tabular}
          {|| c | c ||} 
         \hline
         Notation & Description \\ [0.5ex] 
         \hline\hline
         M & number of input Feature Maps  \\ 
         \hline
         N & number of output Feature Maps \\
         \hline
         K & Kernel size K$\times$K \\ 
         \hline
         L & Size of input feature map L$\times$L \\
         \hline
         $N_{mov}$ & Size of output feature map $N_{mov}$ $\times$ $N_{mov}$ ($N_{mov}=L-K+1$)\\
         \hline
         $B_{IO}$ & bits fetched from SRAM to processor per bank\\
         \hline
         $B_{W}$ & bit width of the weight stored in SRAM\\
         \hline
         $N_{col}$ & number of columns in SRAM array\\ 
         \hline
         $N_{bank}$ & number of SRAM banks\\
         \hline
         $N_{mult}$ & number of multipliers in processor\\
         \hline
         \hline
         $T_{read}$ & time required to fetch data from SRAM to processor\\
         \hline
         $T_{mult}$ & time required to perform multiplication in processor\\
         \hline
         $T_{amac}$ & time required for 1 in memory analog multiply and accumulation \\
         \hline
         $T_{adc}$ & time required for adc operation\\
         \hline
         $E_{read}$ & energy required to fetch data from SRAM to processor\\
         \hline
         $E_{mult}$ & energy required to perform multiplication in processor\\
         \hline
         $E_{amac}$ & energy required for 1 in memory analog multiply and accumulation\\
         \hline
         $E_{adc}$ & energy required for adc operation\\
         \hline
         R & \ number of MAC performed in analog domain\\
         \hline
         $P_{leak}$ & Standby power consumption of SRAM memory\\
         \hline
        \end{tabular}
        }
\end{table}
}

\begin{equation}\label{eq:time}
  \begin{aligned}
        T_{imac}\approx \bigg[\frac{MNK^2}{(N_{col}/B_{W})N_{bank}}\bigg]N_{mov}^2[T_{amac}+T_{adc}/R]
    \end{aligned}
\end{equation}

\begin{equation}\label{eq:energy}
  \begin{aligned}
        E_{imac}\approx MNK^2N_{mov}^2[E_{amac}+E_{adc}/R]
        + P_{leak}T_{in-memory}
  \end{aligned}
\end{equation}

{\centering\begin{table}[t]
    \caption{Parameter Values} 
    \label{tab:parameter}
    \centering
         \begin{tabular}
          {||c|c||c|c||} 
         \hline
         Parameter & Value & Parameter & Value \\ [0.5ex] 
         \hline\hline
         $B_{IO}$ & 16-256  & $B_{W}$ & 5  \\
         \hline
         $N_{bank}$ & 4 & $N_{col}$/$N_{row}$ & 256 \\
         \hline
         $N_{mult}$ & 175 &  & \\
         \hline
         \hline
         $E_{read}$ & \textcolor{black}{\unit[5.2]{pJ}} & $T_{read}$ & \textcolor{black}{\unit[4]{ns}} \\
         \hline
         $E_{mult}$ & \textcolor{black}{\unit[0.9]{pJ}} & $T_{mult}$ & \textcolor{black}{\unit[4]{ns}} \\
         \hline
         $E_{amac}$ & \textcolor{black}{\unit[0.254]{pJ}} & $T_{amac}$ & \textcolor{black}{\unit[1]{ns}} \\
         \hline
         $E_{adc}$ & \textcolor{black}{\unit[0.253]{pJ}} & $T_{adc}$ & \textcolor{black}{\unit[5]{ns}}  \\
         \hline
         $P_{leak}$ & \textcolor{black}{\unit[2.4]{nW}} & R & 10\\
         \hline
        \end{tabular}
\end{table}
}

{\centering \begin{table}[htbp]
\caption{\textcolor{black}{Area estimations}}
        \label{table:area}
        \centering
        \begin{tabular}{||c|c||}
            \hline
            \multicolumn{1}{||p{2cm}|}{\centering Component }  &\multicolumn{1}{|p{3cm}||}{\centering Total area per array \\(\unit[]{$\mu$m\textsuperscript{2}}) }   \\
             \hline
             \hline
             SRAM cell  & 83100\\
             \hline
             ADC &  40800 \\
             \hline
             Accumulator &   30600 \\
             \hline
             DAC &  400 \\
             \hline
             MUX & 2100  \\
             \hline
             Decoder &  4800 \\
             \hline
             Column circuit &  44000 \\
             \hline
             \hline
        \end{tabular}
        
    \end{table}}
    
For a fair comparison, all array parameters are kept the same as \cite{shanbhag-arch} and delay and energy parameters are obtained from circuit simulations. 
The parameters used are presented in Table \ref{tab:parameter}.
Fig. \ref{fig:system} compares the energy, delay and the Energy delay product of this work against baseline. From Fig. \ref{fig:system}, the total improvements in energy, delay and energy-delay product are seen to be 9.42$\times$, 6.24$\times$ and 58.79$\times$ respectively.
Since the array architecture for this work is kept similar to \cite{shanbhag-arch}, we compare the Energy and Delay of this work and find this work to be 2.27$\times$ and 4.83$\times$ better than \cite{shanbhag-arch}.
Table \ref{tab:compare} compares this work with other similar in-memory computing works. 
The plot show in Fig. \ref{fig:EDP_compare} compare the EDP improvements of this work against von Neumann for different $B_{IO}$. 
EDP of this work is 22$\times$ better than von Neumann, even when $B_{IO}$ is made 256.
\textcolor{black}{The area estimates for individual blocks are presented in the Table \ref{table:area}. It can be seen that $\sim$36\% of the total array area would be occupied by the peripherals introduced in this work.}

\section{Conclusion}\label{sec:Conclucion}
In this work we present an in-memory dot product computing primitive using standard 6T SRAM arrays.
We encode the input ($V_{in}$) signal as an analog voltage on the wordline, while the bit significance of the stored input (W) is encoded using precharge pulse.
We perform analog accumulation to reduce the use of energy expensive ADC.
We also perform detailed circuit analysis including random transistor variations to study the effects of such non-idealities on application accuracy.
We develop a circuit-software co-simulation framework including circuit non-idealities and show an accuracy of 88.83\% and 99.05\% on CIFAR-10 and MNIST datasets, respectively. 
Additionally, we evaluate the proposed in-memory compute primitive to compare the system delay/energy with state-of-the-art neural network in-memory accelerators. 
The proposed system is 58.78$\times$ and 8$\times$ better in EDP than the standard von Neumann system and recent neural network accelerator, respectively.


\bibliography{ref1.bib}
\bibliographystyle{IEEEtran}

\begin{IEEEbiography}[{\includegraphics[width=1in,height=1.25in,clip,keepaspectratio]{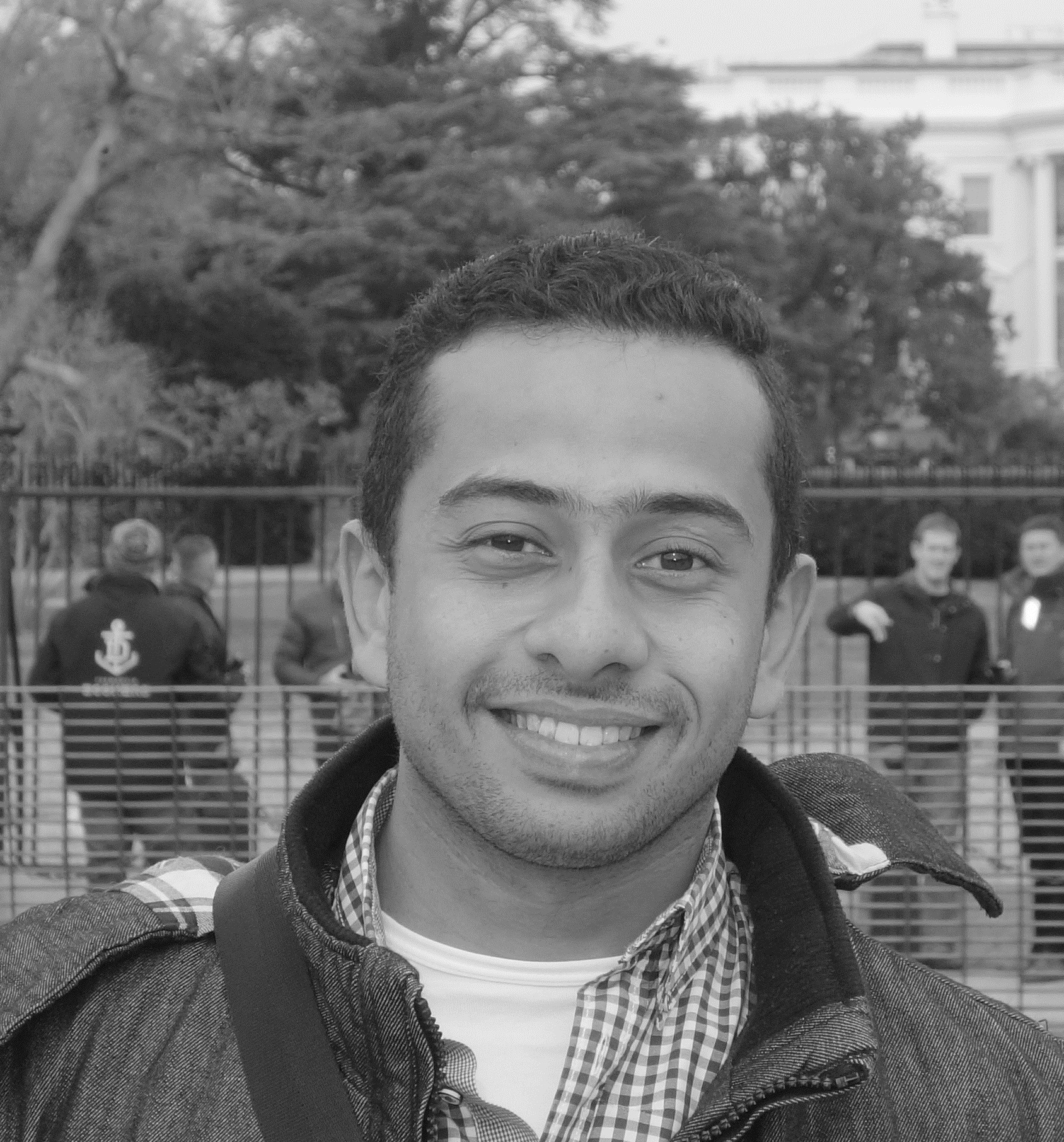}}]{Mustafa Ali} received his B.Sc. ans M.Sc. degrees in Electrical Engineering from MTC, Cairo, Egypt in 2011, 2016 respectively. He achieved the 1st rank in undergraduate in MTC, 2011. He was honored the Duty Medal for excellent performance during his studies. He worked on flexible electronics applications using TFTs in his M.Sc from 2014 to 2016. Additionally, he worked as a TA and RA at MTC from 2013 to 2017. Mustafa was also a hardware and embedded systems engineer at Integreight, Inc. from 2012 to 2017. He joined the Nano-electronics Research Lab (NRL), Purdue University in Spring 2018 and he is currently pursuing his Ph.D. under the guidance of Prof. Roy. His research interest lies in accelerating Brain-inspired Computing and machine learning. He is also interested in in-memory computing based on CMOS and post-CMOS devices.
\end{IEEEbiography}
\begin{IEEEbiography}[{\includegraphics[width=1in,height=1.25in,clip,keepaspectratio]{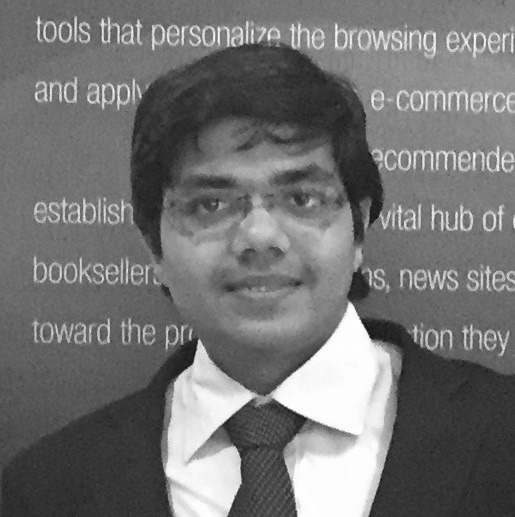}}]{Akhilesh Jaiswal}
 received the B.Tech. degree from Shri Guru Gobind Singhji Institute of Engineering and Technology, Nanded, India, in 2011, the M.S. degree in electrical engineering from the University of Minnesota, Minneapolis, MN, USA, in 2014 and PhD in electircal and communication engineering from Purdue University, West Lafayete, USA, in 2019.
He joined the Globalfoundries Laboratory, Malta, NY, USA in 2019. He was an Intern at the Globalfoundries Laboratory, Malta, NY, USA, in 2017. His current research interests include modeling and simulation of spin devices for on-chip memory/logic/nueromorphic applications, and
neuromorphic applications using analog and digital CMOS circuits. Besides he is also interested in in-memory computing based on emerging and CMOS devices.
\end{IEEEbiography}
 \begin{IEEEbiography}[{\includegraphics[width=1in,height=1.25in,clip,keepaspectratio]{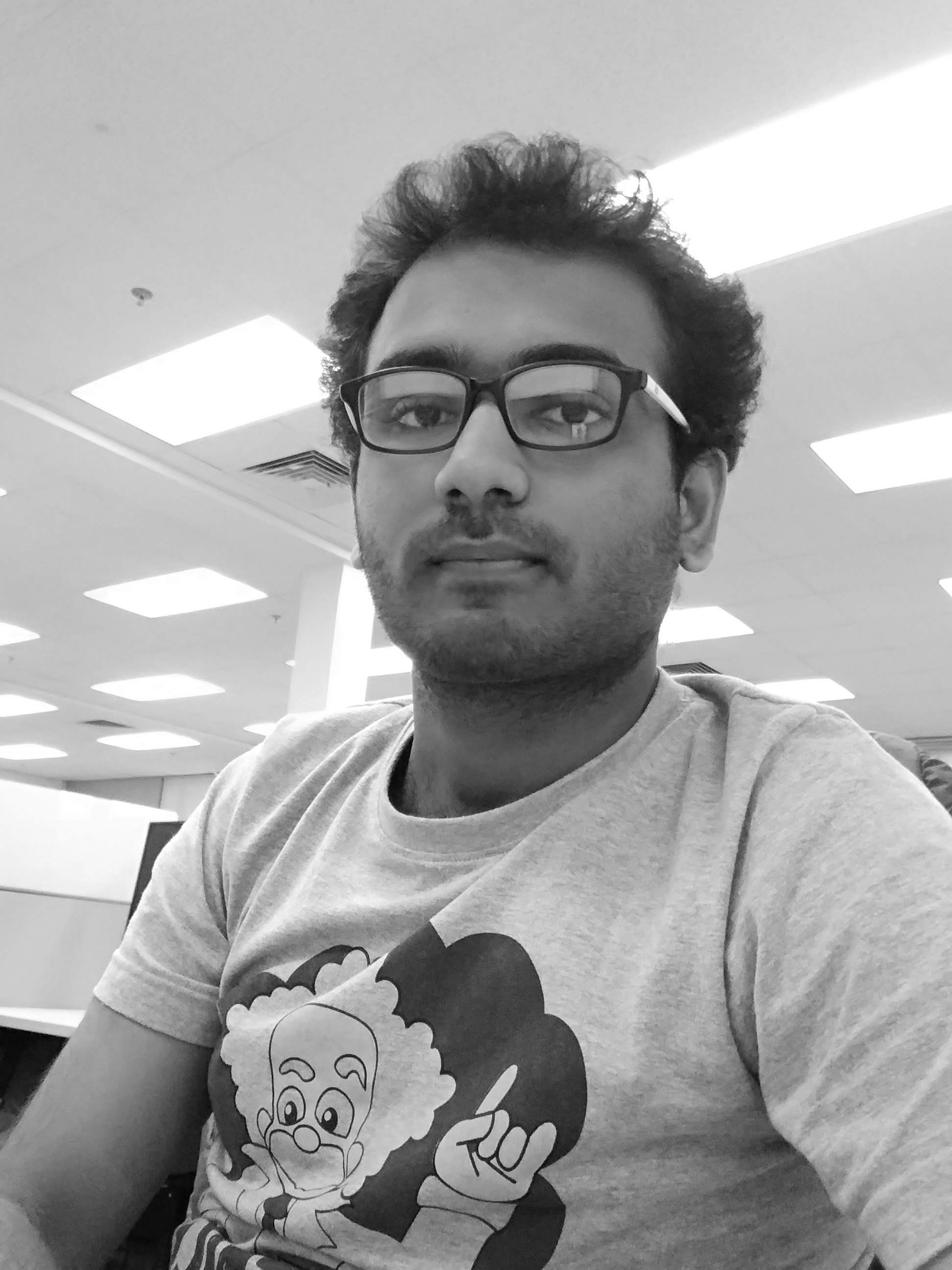}}]{Sangamesh Kodge} received the B.Tech. degree in electrical engineering  from  IIT  Kharagpur,  Kharagpur,  India ,  in  2018. During his tenure at IIT Kharagpur, he has received Institute Silver medal, Rajender Kumar Khanna Memorial prize on being adjudged to be the best student in the order of merit and System Society Award for the best project on systems. He was a research intern and NTU-India Connect Fellow at Nanyang Technological University (NTU), Singapore in 2017. He joined Nanoelectronics research lab at Purdue in fall 2018 and is currently pursuing his PhD under the guidance of Prof. Kaushik Roy. His primary research interest lie in brain-inspired computing, explanable machine learning and theoretical machine learning.
 \end{IEEEbiography} 
\begin{IEEEbiography}[{\includegraphics[width=1in,height=1.25in,clip,keepaspectratio]{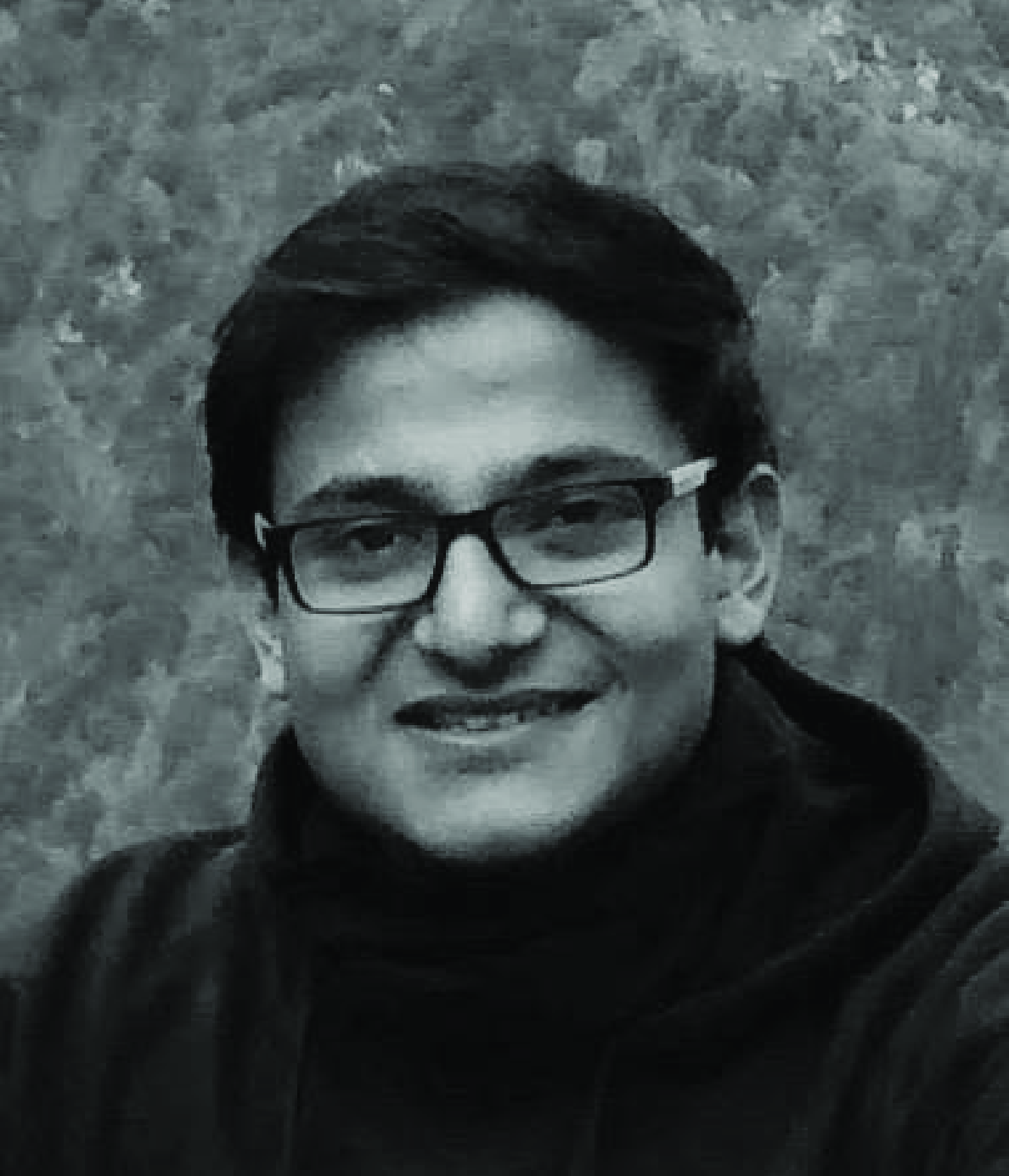}}]{Amogh Agrawal}
received the B.Tech degree in electrical engineering from the Indian Institute of Technology, Ropar, India in 2016. He was a research intern at University of Ulm, Germany in 2015, under the DAAD (German Academic Exchange Service) Fellowship. He joined the Nanoelectronics research lab in 2016, and is currently pursuing his PhD degree at Purdue University under the guidance of Prof. Kaushik Roy. He was a Technology Development Intern at GLOBALFOUNDRIES, Malta, NY, USA, during the summer of 2018.His primary research interests include modeling and simulation of spin devices for application in logic, memories and neuromorphic computing. He is also looking at in-memory computing techniques for neuromorphic computing using CMOS and beyond-CMOS memories. He was awarded the Directors Gold Medal for his all-round performance, and University Silver Medal for his academic achievements at IIT Ropar. He is a recipient of the Andrews Fellowship from Purdue University, in 2016.
\end{IEEEbiography}
\begin{IEEEbiography}[{\includegraphics[width=1in,height=1.25in,clip,keepaspectratio]{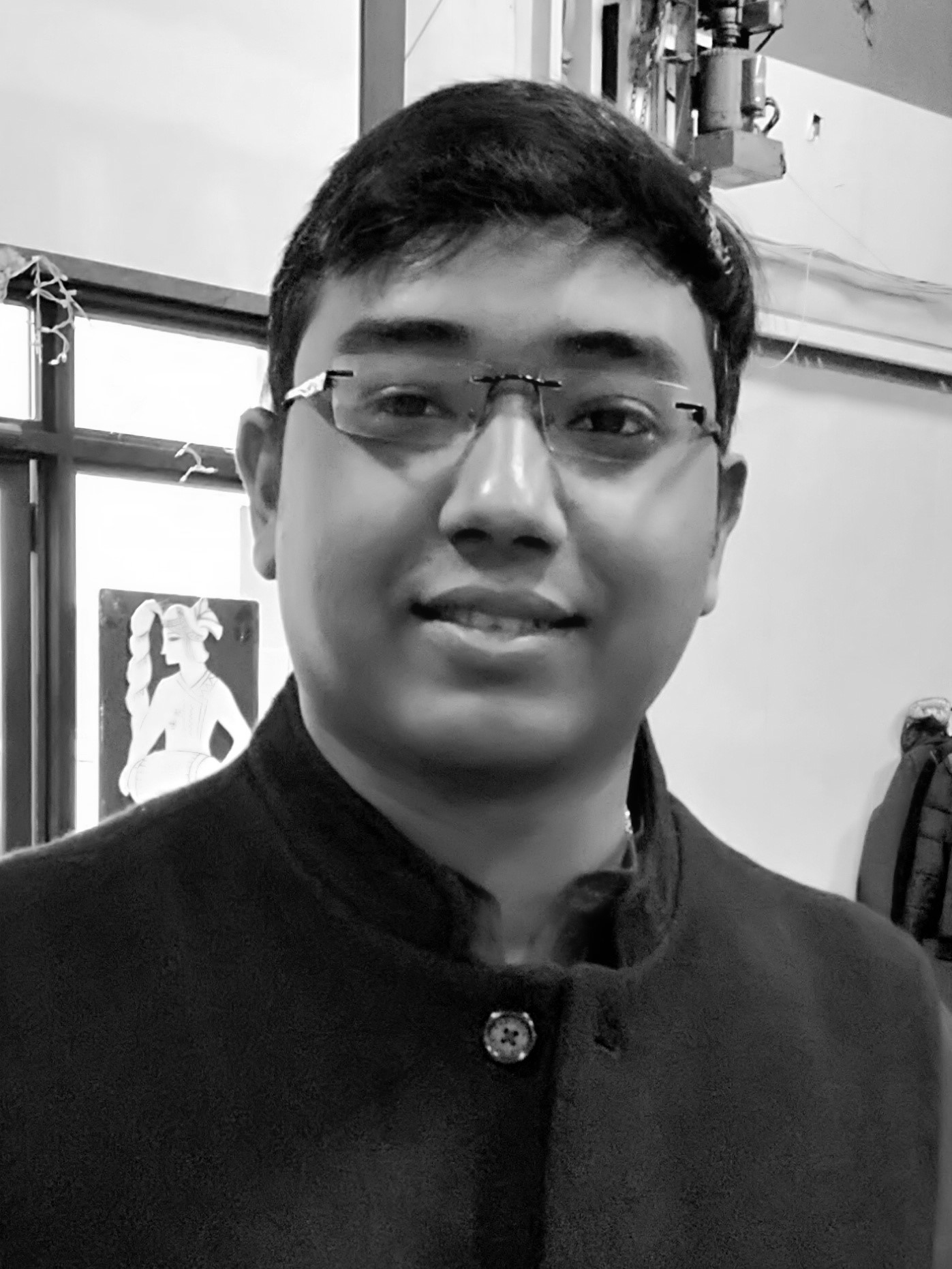}}]{Indranil Chakraborty} received the B.Engg. degree in Electronics and Telecommunication Engineering from Jadavpur University, Kolkata, India, in 2013, and the M.Tech. degree in Electrical Engineering from Indian Institute of Technology Bombay, Mumbai, India, in 2016.  He was the recipient of best M. Tech thesis award and academic excellence award during his time at IIT Bombay for his academic performance. Since Fall 2016, he has been pursuing the Ph.D. degree with the Nanoelectronics Research Laboratory, Purdue University, West Lafayette, IN, USA. His primary research interests lie in hardware for machine-learning using CMOS and emerging technologies.
\end{IEEEbiography}
\begin{IEEEbiography}[{\includegraphics[width=1in,height=1.25in,clip,keepaspectratio]{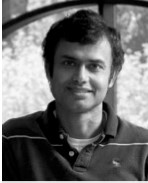}}]{Kaushik Roy} received the B.Tech. degree in electronics and electrical communications engineering  from  IIT  Kharagpur,  Kharagpur,  India  and the  Ph.D.  degree  from  the  Department  of  Electrical  and  Computer  Engineering,  University  of  Illinois  at  Urbana-Champaign,  Champaign,  IL,  USA,in 1990. He   was   with   the   Semiconductor   Process   and Design Center, Texas Instruments, Dallas, TX, USA, where  he  was  involved  in  field-programmable  gate array  architecture  development  and  low-power  circuit  design.  He  was  a  Faculty  Scholar  at  Purdue  University,  West Lafayette, IN, USA, from  1998  to  2003.  He was  a Research  Visionary  Board  Member of Motorola Labs in 2002. He held the M. K. Gandhi Distinguished  Visiting Faculty, IIT Bombay, Mumbai, India. He joined  the Electrical  and Computer Engineering  Faculty,  Purdue  University,  West  Lafayette,  IN,  USA,  in  1993,where  he  is  currently  an  Edward G. Tiedemann  Jr.  Distinguished  Professor. He  has  authored   over  600  papers  in  refereed   journals   and  conferences, holds  15  patents,  graduated  60  Ph.D.  students,  and  is  a  coauthor  of  two books: Low-Power CMOS VLSI Circuit Design (New York, NY, USA: Wiley, 2009)  and Low  Voltage,   Low  Power  VLSI  Subsystems (New  York,  NY, USA: McGraw-Hill, 2005). His current research interests include spintronics, device circuit codesign for nanoscale silicon and nonsilicon technologies, low-power  electronics  for  portable  computing  and  wireless  communications,  and new computing  models enabled  by emerging  technologies. Dr. Roy received the U.S. National Science Foundation Career Development Award  in  1995,  the  IBM  Faculty  Partnership  Award,  the  ATT/Lucent  Foundation Award, the 2005 SRC Technical Excellence Award, the SRC Inventors Award,   the  Purdue  College   of  Engineering   Research   Excellence   Award, the Humboldt Research Award in 2010, the 2010 IEEE Circuits and Systems Society  Technical  Achievement  Award,  the  Distinguished  Alumnus  Award from IIT Kharagpur, the Fulbright-Nehru Distinguished Chair, and best paper awards at the 1997 International Test Conference, the IEEE 2000 International Symposium  on  Quality  of  IC  Design,  the  2003  IEEE  Latin  American  Test Workshop,  the  2003  IEEE  Nano,  the  2004  IEEE  International  Conference on Computer Design, the 2006 IEEE/ACM International Symposium on Low Power Electronics \& Design, and the 2005 IEEE Circuits and System Society Outstanding  Young  Author  Award (Chris  Kim), the 2006 IEEE Transactions on  VLSI  Systems  Best  Paper  Award,  the  2012  ACM/IEEE  International Symposium   on   Low   Power   Electronics   and   Design   Best   Paper   Award,the  2013  IEEE  Transactions  on  VLSI  Best  Paper  Award.  He  has  been  on the Editorial  Board  of  IEEE  DESIGN AND TEST, the IEEE  TRANSACTIONS ON CIRCUITS AND SYSTEMS, the  IEEE  TRANSACTIONS ON VERY LARGE SCALE INTEGRATION(VLSI)  SYSTEMS,  and  the  IEEE  TRANSACTIONS ON ELECTRON DEVICES.  He  was  the  Guest  Editor  for  the  Special  Issue on  Low-Power  VLSI  in  IEEE  DESIGN  ANDTESTin  1994  and  the  IEEE TRANSACTIONS ON VERY LARGE SCALE INTEGRATION (VLSI)  SYSTEMS in 2000, the IEE Proceedings-Computers and Digital Techniques in 2002, and the  IEEE  JOURNAL  ON EMERGING AND SELECTED TOPICS IN CIRCUITS AND SYSTEMS in 2011.
\end{IEEEbiography}

\end{document}